\documentclass{aa}  
\usepackage{natbib}
\usepackage{graphicx}
\usepackage{longtable}
\usepackage{txfonts}
\usepackage{lscape}
\begin{document} 
%
  \title{The First Simultaneous 3.5 and 1.3\,mm Polarimetric Survey of Active Galactic Nuclei in the Northern Sky}

   \author{I. Agudo
               \inst{1,2,3}
          \and
               C. Thum
	       \inst{4}
          \and
               J. L. G\'omez
	       \inst{1}
         \and
	       H. Wiesemeyer
	       \inst{5}
             }
           
  \institute{Instituto de Astrof\'{\i}sica de Andaluc\'{\i}a (CSIC),  
                 Apartado 3004, E-18080 Granada, Spain
         \and
                Institute for Astrophysical Research, Boston University, 
                725 Commonwealth Avenue, Boston, MA 02215, USA
         \and
                Current Address: Joint Institute for VLBI in Europe, 
                Postbus 2, NL-7990 AA Dwingeloo, the Netherlands, \email{agudo@jive.nl}
         \and
                Instituto de Radio Astronom\'ia Millim\'etrica, 
                Avenida Divina Pastora, 7, Local 20, E--18012 Granada, Spain
         \and
                Max-Planck-Institut f\"ur Radioastronomie, 
                Auf dem H\"ugel, 69, D-53121, Bonn, Germany
             }


  \abstract
   {Short millimeter observations of radio-loud active galactic nuclei (AGN) offer an excellent opportunity to study the physics of their synchrotron-emitting relativistic jets, from where the bulk of radio and millimeter emission is radiated.
   On one hand, AGN jets and their emission cores are significantly less affected by Faraday rotation and depolarization than at longer wavelengths.
   On the other hand, the millimeter emission of AGN is dominated by the compact innermost regions in the jets, where the jet can not be seen at longer wavelengths due to synchrotron opacity.}
   {We present the first dual frequency simultaneous 86\,GHz and 229\,GHz polarimetric survey of all four Stokes parameters of a large sample of 211 radio loud active galactic nuclei, designed to be flux limited at 1\,Jy at 86\,GHz.}
   {The observations were most of them made in mid August 2010 using the XPOL polarimeter on the IRAM 30\,m millimeter radio telescope.}
   {Linear polarization detections above $3\sigma$ median level of $\sim1.0$\,\% are reported for 183 sources at 86\,GHz, and for 23 sources at 229\,GHz, where the median $3\sigma$ level is $\sim6.0$\,\%.
     We show a clear excess of the linear polarization degree detected at 229\,GHz with regard to that at 86\,GHz by a factor of $\sim1.6$, thus implying a progressively better ordered magnetic field for blazar jet regions located progressively upstream in the jet.
     We show that the linear polarization angle, both at 86 and 229\,GHz, and the jet structural position angle for both quasars and BL~Lacs do not show a clear preference to align in either parallel or perpendicular directions. 
     Our variability study with regard to the 86\,GHz data from our previous survey points out a large degree variation of total flux and linear polarization in time scales of years by median factors of $\sim1.5$ in total flux, and $\sim1.7$  in linear polarization degree --maximum variations by factors up to $6.3$, and $\sim5$, respectively--, with 86\,\% of sources showing linear polarization angles evenly distributed with regard to our previous measurements.}
   {}

   \keywords{Galaxies: active
   - galaxies: jets
   - quasars: general 
   - BL~Lacertae objects: general
   - polarization
   - surveys}

   \maketitle

\section{Introduction}
\label{Intr}

Among all classes of active galactic nuclei (AGN), radio-loud AGN are characterized by having pairs of relativistic jets of highly energized, magnetized plasma that are ejected along the rotational poles of the SMBH--disk system \citep[e.g.,][]{Blandford:1982p6283}. 
In particular, blazars, the most exotic class of radio-loud AGN, stand out by wild variability of their non-thermal jet-emission from radio to $\gamma$-rays. 
Members of this class include BL~Lacs, and flat spectrum radio quasars (FSRQ, the high power version of the former). 
The remarkable properties of blazars are affected by strong relativistic effects that beam their radiation and shorten the variability time scales.
These properties include superluminal motions \cite[e.g.,][]{Gomez:2001p201, Jorstad:2005p264, Agudo:2012p17361}, Doppler boosted emission of the jet pointing at a small angle to the observer \citep[e.g.,][]{Kadler:2004p5629}, substantial changes in flux and linear polarization, sometimes correlated on several spectral ranges, on time scales from months \citep[e.g.,][]{Bach:2006p354,Agudo:2011p14707,Agudo:2011p15946} to hours \citep[e.g.,][]{Ackermann:2010p13506, Tavecchio:2010p14858}, as well as changes of jet structure \cite[e.g.,][]{2007AJ.134.799J, Agudo:2007p132}, and polarized synchrotron and inverse-Compton emission all along the spectrum \cite[e.g.,][]{Marscher:2010p11374,Jorstad:2010p11830,Agudo:2011p14707,Agudo:2011p15946,Wehrle:2012p17706}.
AGN with powerful jets oriented at larger angles to the line of sight (i.e. radio galaxies, the remaining prominent class of radio-loud AGN) suffer less from such relativistic effects, therefore displaying smaller apparent luminosities and longer time scales of variability, although they are thought to be the same astrophysical objects as BL~Lacs and FSRQ, the latter being the better oriented counterparts of the former \citep{Urry:1995p6301}.

Explaining the existence and physical properties of jets from AGN remains one of the greatest current challenges in high-energy astrophysics. 
Observations of radio-loud AGN at short millimeter wavelengths offer an excellent opportunity with regard to radio centimeter observations to study these objects for several reasons.
First, like AGN radio emission, millimeter emission is well known to be radiated predominantly from the jet, with essentially no contribution from the host AGN or its surroundings, which facilitates the interpretation of the measurements in terms of jet physics.
Second, AGN jets, and in particular their millimeter emitting cores, are significantly less affected by Faraday rotation and depolarization than at longer wavelengths \citep[see e.g.][]{Zavala:2004p138,Agudo:2010p12104}.
Third, the millimeter emission of AGN is dominated by the compact innermost regions in the jets \citep[e.g.][]{2007AJ.134.799J,Lee:2008p301}, where the jet can not be seen at longer wavelengths due to synchrotron opacity.  

In 2005, we used the XPOL polarimeter \citep{2008PASP..120..777T} on the IRAM 30\,m Telescope to perform the first 3.5\,mm (86\,GHz) polarization survey of radio loud AGN over a large (145 source) sample \citep{Agudo:2010p12104}. 
Within the available data, we detected linear polarization above $3\sigma$ levels of $\sim1.5$\,\% for 76\,\% of the sample.
Our results pointed out an excess, by a factor of $\sim2$, of 86\,GHz linear polarization degree with regard to that at radio wavelengths (15\,GHz), suggesting that either the region of bulk millimeter emission has a better ordered magnetic field, or that the radio emission is strongly Faraday depolarized with regard to that at millimeter wavelengths.
We also reported a trend of decreasing luminosity towards larger linear polarization degrees within our entire sample, perhaps indicating lower magnetic field order for larger luminosity jets.
Moreover, in contrast to what it was found at radio wavelengths \citep[see][ and references therein]{Lister:2005p261}, we do not find a relation between the linear polarization angle and the jet structural position angle in either quasars of BL~Lacs. 
Confirmation of this result would imply that no AGN class matches the conditions to show linear polarization angles either parallel or perpendicular to the jet, i.e. that they are intrinsically non-axisymmetric.

Aiming to improve the statistical confidence with regard to previous studies, we performed a new simultaneous 3.5 (86\,GHz) and 1.3\,mm (229\,GHz) AGN polarimetric survey with the XPOL polarimeter on the IRAM 30\,m Telescope in polarimetric mode over an updated and increased sample of 211 sources that was designed to be 1\,Jy total-flux limited.
Our new observations were also contemporaneous with current large surveys at other spectral ranges, therefore allowing for multi spectral range studies.
The dual frequency configuration of our observing program also allowed us to search for large Faraday rotation effects in bright AGN, whereas the comparison with our previous survey allowed us to study the total flux and polarization AGN variability at short millimeter wavelengths in bright AGN.
In this paper we describe this new survey and the results from the analysis of the resulting data.
The comparison of the data from this new survey with $\gamma$-ray measurements from the Fermi-LAT 2-year AGN Catalog \citep{Ackermann:2011p17255}, as well as a detailed statistical study of their correlation are presented in a separate paper (Agudo et al., A\&A, in prep).
 
\section{The sample}
\label{Samp}

Table~\ref{T1} shows the list of 211 sources covered by our new survey together with some of the most prominent properties of every source.
This list comprises the 102 AGN in our previous 3.5 mm AGN Polarization survey brighter than 0.9\,Jy plus 109 additional AGN sources with estimated\footnote{Note that the $S_{\rm{90GHz}}$ WMAP fluxes were estimated from extrapolation from the WMAP spectra and spectrum $S_{\rm{60GHz}}$ where a $S_{\rm{90GHz}}$ was not available.} 90\,GHz flux density $S_{\rm{90GHz}}>0.9$\,Jy contained in the WMAP (Wilkinson Microwave Anisotropy Probe) 7-Year Catalog of Point Sources \citep{Gold:2011p18071} with declination $>-30^{\circ}$ in J2000.0.

Hence, our source list was designed to build a 3.5\,mm 1\,Jy flux limited complete sample of AGN. 
However, our observations have revealed that among the 211 observed sources, 108 of them showed $S_{\rm{86}}<1$\,Jy, and even 47 of them had $S_{\rm{86}}<0.5$\,Jy.
We attribute this result both to total-flux source variability --that reached a factor 1.5 for 50\,\% of sources in our sample (see Section~\ref{Var})--, and also to inaccuracies in the estimated $S_{\rm{90GHz}}$ fluxes from the WMAP 7-Year Catalog for candidate sources.
Indeed, $~\sim40$ sources of the WMAP 7-year sample do not match our selection criteria, while they do so in the 9-year WMAP sample which is assumed to include more precise measurements of the weaker objects.

Radio-loud AGN with relatively large luminosities at millimeter wavelengths are dominated by relativistically enhanced flat spectrum radio emission from the innermost jet regions in AGN, whereas the emission from steep spectrum radio sources --typically weak or undetectable at millimeter wavelengths-- comes from giant radio lobes of radio galaxies.
Also, because of the spectral criteria applied to select the sources in our previous 3.5\,mm survey, its sample was dominated by flat spectrum radio sources (i.e. FSRQ and BL~Lacs).
Therefore, our sample is dominated by blazars.
In particular, our new 3.5 \& 1.3\,mm AGN Polarization Survey sample contains 152 quasars,  32 BL~Lacs, and  21 radio galaxies, with 6 unclassified sources, i.e., not contained in the \citet{VeronCetty:2006p4900} catalog.
Thus, our new 3.5 \& 1.3\,mm AGN Polarization Survey sample is adequate for studies of the mm polarimetric properties of quasar and BL~Lac blazars.

Also, 110 of our sources are contained in the MOJAVE 15\,GHz VLBI Survey of bright AGN jets \cite{Lister:2009p5316}.
83\,\% of MOJAVE sources are also in our sample. 
This high rate of coincidence shows that the two samples are adequate for comparative studies.  

The source redshift in the our entire sample ranges from $z=0.00068$ to $z=3.408$, with mean and median at  $\bar{z}=0.937$ and $\tilde{z}=0.859$, respectively.

\begin{table*}
\caption{\label{T1} Source properties (truncated table).}
\begin{tabular}{ccccccccccc}
\hline\hline
Source name        &          &           &         &     &     &  $V$ &  $\phi_{\rm{jet}}$ & $\nu_{\rm{obs}}$ &  Ref. & \emph{Fermi}  \\
  (IAU)             &    Alias & (J2000.0) &  (J2000.0) & z & Cl.  &  mag & [$^{\circ}$] & [GHz] & ($\phi_{\rm{jet}}$)  & 2LAC? \\
(1) &  (2) &  (3) &  (4) &  (5) &  (6) &  (7) &  (8)  & (9) &  (10)  & (11)\\
\hline
\object{0003+380}            & \object{           S4 0003+38} & 00 05 57.1754 & +38 20 15.148 & 0.229 & G & 19.9 & 106 &  15 &   9 &  Y  \\ 
\object{0003-066}            & \object{             NRAO   5} & 00 06 13.8928 & -06 23 35.334 & 0.347 & B & 18.5 & 181 &  86 &   6 & ... \\ 
\object{0007+106}            & \object{            III Zw  2} & 00 10 31.0058 & +10 58 29.504 & 0.089 & G & 15.4 & 221 &  86 &   6 & ... \\ 
\object{0017+200}            & \object{         PKS 0017+200} & 00 19 37.8545 & +20 21 45.644 &  ...  & B & 20.6 & 305 &  43 &  10 & ... \\ 
\object{0027+056}            & \object{         PKS 0027+056} & 00 29 45.8962 & +05 54 40.712 & 1.317 & Q & 15.9 & 163 &  15 &   9 & ... \\ 
\object{0035-252}            & \object{         PKS 0035-252} & 00 38 14.7354 & -24 59 02.235 & 1.196 & Q & 20.7 & 316 &   8 &  10 &  Y  \\ 
\object{0045-255}            & \object{             NGC  253} &  00 47 33.120 &  -25 17 17.59 & 0.001 & G & 14.0 & ... & ... & ... &  Y  \\ 
\object{0048-097}\tablefootmark{a} & \object{          PKS 0048-09} & 00 50 41.3173 & -09 29 05.209 & 0.635 & B & 17.4 &   7 &  86 &   6 &  Y  \\ 
\object{0048-071}            & \object{         PKS 0048-071} & 00 51 08.2098 & -06 50 02.228 & 1.975 & Q & 19.5 & ... & ... & ... &  Y  \\ 
\object{0055+300}            & \object{             NGC 0315} & 00 57 48.8833 & +30 21 08.812 & 0.016 & G & 12.5 & 308 &  15 &   5 & ... \\ 
\object{0059+581}\tablefootmark{b} & \object{         TXS 0059+581} & 01 02 45.7623 & +58 24 11.136 & 0.644 & Q & 17.3 & 235 &  15 &   8 &  y  \\ 
\object{0106+013}            & \object{          PKS 0106+01} & 01 08 38.7710 & +01 35 00.317 & 2.107 & Q & 18.4 & 235 &  15 &   8 &  Y  \\ 
\object{0112-017}            & \object{         PKS 0112-017} & 01 15 17.0999 & -01 27 04.576 & 1.365 & Q & 17.5 & 118 &  15 &   5 & ... \\ 
\object{0113-118}            & \object{          PKS 0113-11} & 01 16 12.5219 & -11 36 15.432 & 0.672 & Q & 19.0 & 338 &  15 &   9 &  Y  \\ 
\hline
\end{tabular}
\tablefoot{Columns are as follows: (1) IAU B1950.0 source name, (2) source name in other catalogs, (3) and (4) J2000.0 observing right ascension and declination, (5) redshift from \citet{VeronCetty:2006p4900}, (6) optical classification from  \citet{VeronCetty:2006p4900}, (7) V magnitude from  \citet{VeronCetty:2006p4900}, (8) jet position angle ($\phi_{\rm{jet}}$), (9) VLBI observing frequency for the image from which $\phi_{\rm{jet}}$ was obtained, (10) References from which $\phi_{\rm{jet}}$ was obtained (see below), (11) \emph{Fermi}-2LAC source (Y=yes, y=yes but not in the clean list).}
\tablebib{(1) Agudo et al. (2007); (2) BU- Blazar Group Web Page; (3) Fomalont et al. (2000); (4) Jorstad et al. (2005); (5) Kellermann et al. (2004); (6) Lee et al. (2008); (7) Lister et al. (2001); (8) Lister \& Homan (2005); (9) MOJAVE Web Page; (10) USNO Radio Reference Frame Image Database Web Page;  (11) Xu et al. (1995).}
\tablefoottext{a}{Redshift from \cite{Ackermann:2011p17255}.}
\tablefoottext{b}{$V$-mag$=R$-mag from NED. Not present in \citet{VeronCetty:2006p4900} catalog. Redshift from \citet{SowardsEmmerd:2005p4877}.}
\end{table*}

\section{Observations and data reduction}
\label{Obs}

The observations were performed with the XPOL polarimeter \citep{2008PASP..120..777T} connected to the E090 and E230 EMIR receiver system \citep{Carter:2012p18089} on the IRAM 30\,m Telescope.
The advanced design of EMIR allowed to perform the observations simultaneously at 3.5\,mm (86\,GHz, with the E090 pair of orthogonal receivers) and 1.3\,mm (229\,GHz, with the E230 pair of orthogonal receivers) without loss of significant signal at either waveband \citep{2008PASP..120..777T}.
The flexibility of the VESPA autocorrelator in polarimetric mode (i.e. XPOL) also allows to simultaneously process the 86 and 229\,GHz signals, although with a limited bandwidth of 640\,MHz for each of the two orthogonally polarized receivers of the E090 band, and of 320\,MHz for those of the E230 band.

The bulk of the observing program was performed from 13 to 16, August 2010, although a small number of remaining observations were done several weeks before (from May 5, 2010) or after (until June 16,2011) to complete observations of missing sources or initially unreliable measurements, see Table~\ref{T2}.

To perform the observations, we employed the standard XPOL set--up and calibration method discussed in \citet{2008PASP..120..777T}.
Our observing strategy was essentially the same as the one used for our previous 3.5\,mm AGN Polarimetric Survey \citep{Agudo:2010p12104}.  
We made a continuous refinement of the pointing model of the telescope before every polarization integration, as well as frequent measurements of the focus parameters and of polarization calibrators (Mars and Uranus).

For the data reduction we followed the procedure in \citet{Agudo:2010p12104}.
The instrumental polarization removed from the data was the one estimated from observations of unpolarized calibrators, i.e. $Q_{i,\rm{86}}=-0.7\pm0.3$\,\%, $U_{i,\rm{86}}=-0.3\pm0.2$\,\%, and $V_{i,\rm{86}}=0.0\pm0.3$\,\% for 86\,GHz observations, and $Q_{i,\rm{229}}=-0.4\pm1.5$\,\%, $U_{i,\rm{229}}=-1.1\pm1.5$\,\%, and $V_{i,\rm{229}}=2.4\pm0.6$\,\% for 229\,GHz.
These instrumental polarization parameters are fully consistent with those measured during the first months of operation of XPOL connected to EMIR.
After applying all polarization calibrations, we obtained final polarization-error medians of $\Delta \tilde{m}_{\rm{L}}^{\rm{86}}\approx$0.32\,\%,  $\Delta \tilde{\chi}^{\rm{86}}\approx$3\,$^{\circ}$, and  $\Delta \tilde{m}_{\rm{C}}^{\rm{86}}\approx$0.35\,\% for the linear polarization degree  ($m_{\rm{L}}$), the linear polarization electric vector position angle ($\chi$), and the circular polarization degree ($m_{\rm{C}}$) at 86\,GHz, respectively.
For the observations at 229\,GHz we obtained $\Delta \tilde{m}_{\rm{L}}^{\rm{229}}\approx$2\,\%,  $\Delta \tilde{\chi}^{\rm{229}}\approx$7\,$^{\circ}$, and  $\Delta \tilde{m}_{\rm{C}}^{\rm{229}}\approx$2\,\%.
For the conversion from antenna temperatures to flux densities we used standard calibration factors of $C_{\rm{Jy/K}^{\rm{86}}}=6.4$\,Jy/K, and $C_{\rm{Jy/K}^{\rm{229}}}=9.3$\,Jy/K at 86 and 229\,GHz, respectively \citep[e.g.,][]{Agudo:2006p203}.

\begin{table*}
\caption{\label{T2} Summary of observing results (truncated table).}
\begin{tabular}{ccccccccccc}
\hline\hline
Source & MJD  & $t_{\rm{int}}$ & $S_{86}$ & $m_{86,\rm{L}}$ & $\chi_{86}$ & $m_{86,\rm{C}}$  & $S_{229}$ & $m_{229,\rm{L}}$ & $\chi_{229}$ & $|\rm{RM}_{86,229}^{\rm{upper\,limit}}|$ \\
name  & [d] & [min] &  [Jy] & [\%] & [$^{\circ}$] & [\%] &  [Jy] & [\%] & [$^{\circ}$]  & [rad\,$\rm{m^{-2}}$] \\
(1) &  (2) &  (3) &  (4) & (5) &  (6) &  (7) &  (8) & (9) &  (10)  &  (11) \\
\hline
\object{0003+380} & 55422.9971 &   6.6 &  0.78$\pm$0.04 &  4.6$\pm$0.3 & 126.7$\pm$ 2.3 &     $<$ 1.2    &  0.54$\pm$0.03 &     $<$  6.2     &      ...       &        ...          \\ 
\object{0003-066} & 55406.0942 &   6.6 &  1.75$\pm$0.09 & 11.6$\pm$0.3 &  17.0$\pm$ 0.7 &     $<$ 1.0    &  1.20$\pm$0.06 &  16.7$\pm$ 2.4  &  10.4$\pm$ 4.0 &   2.04e+04    \\ 
\object{0007+106} & 55423.0650 &   6.6 &  0.68$\pm$0.03 &  3.1$\pm$0.3 &  19.6$\pm$ 3.1 &     $<$ 1.2    &  0.53$\pm$0.03 &     $<$  6.7     &      ...       &        ...          \\ 
\object{0017+200} & 55423.1097 &   6.6 &  0.29$\pm$0.01 &  4.1$\pm$0.6 &  77.5$\pm$ 3.6 &     $<$ 1.8    &  0.31$\pm$0.02 &     $<$  8.8     &      ...       &        ...          \\ 
\object{0027+056} & 55423.0577 &   6.9 &  0.26$\pm$0.01 &  5.1$\pm$0.7 &  92.3$\pm$ 3.9 &     $<$ 2.0    &  0.20$\pm$0.01 &     $<$ 15.7     &      ...       &        ...          \\ 
\object{0035-252} & 55423.1447 &   6.9 &  0.78$\pm$0.04 &  1.4$\pm$0.4 & 110.0$\pm$ 6.9 &     $<$ 1.2    &  0.43$\pm$0.03 &     $<$  9.9     &      ...       &        ...          \\ 
\object{0045-255} & 55423.1374 &   6.6 &  0.40$\pm$0.02 &     $<$ 1.6   &      ...       &     $<$ 1.7    &  0.79$\pm$0.04 &     $<$  7.5     &      ...       &        ...          \\ 
\object{0048-097} & 55423.2353 &   6.6 &  0.17$\pm$0.01 &  9.4$\pm$1.0 &  88.4$\pm$ 2.8 &     $<$ 3.2    &  0.14$\pm$0.01 &     $<$ 19.9     &      ...       &        ...          \\ 
\object{0048-071} & 55423.2452 &   6.6 &  0.74$\pm$0.04 &  2.6$\pm$0.3 &  72.5$\pm$ 3.6 &     $<$ 1.1    &  0.33$\pm$0.02 &     $<$  9.7     &      ...       &        ...          \\ 
\object{0055+300} & 55423.0269 &   6.6 &  0.51$\pm$0.03 &  1.7$\pm$0.4 & 159.8$\pm$ 6.4 &     $<$ 1.3    &  0.26$\pm$0.02 &     $<$ 11.1     &      ...       &        ...          \\ 
\object{0059+581} & 55355.1482 &   3.3 &  2.12$\pm$0.11 &  1.8$\pm$0.2 &  48.5$\pm$ 5.2 &     $<$ 1.0    &  1.36$\pm$0.07 &     $<$  5.3     &      ...       &        ...          \\ 
\object{0106+013} & 55423.2751 &   3.3 &  2.44$\pm$0.12 &  1.5$\pm$0.3 & 174.8$\pm$ 4.4 &     $<$ 1.0    &  1.52$\pm$0.08 &     $<$  5.2     &      ...       &        ...          \\ 
\object{0112-017} & 55423.2650 &   6.6 &  0.22$\pm$0.01 &  4.5$\pm$0.7 &  82.6$\pm$ 4.8 &     $<$ 2.6    &  0.42$\pm$0.02 &     $<$  7.1     &      ...       &        ...          \\ 
\object{0113-118} & 55423.2552 &   6.6 &  1.24$\pm$0.06 &  3.6$\pm$0.3 &  23.4$\pm$ 2.3 &     $<$ 1.1    &  0.36$\pm$0.02 &     $<$  9.3     &      ...       &        ...          \\ 
\hline
\end{tabular}
\tablefoot{Columns are as follows: (1) IAU B1950.0 source name, (2) MJD observing date, (3) integration time, (4) 86.2\,GHz flux density, (5) 86.2\,GHz fractional linear polarization, (6) 86.2\,GHz linear polarization electric vector position angle, (7) 86.2\,GHz fractional circular polarization, (8) 229\,GHz flux density, (9) 229\,GHz fractional linear polarization, (10) 229\,GHz linear polarization electric vector position angle,  (11) Absolute value of the maximum rotation measure allowed from $\chi_{\rm{86}}$ and $\chi_{\rm{229}}$, and computed as a $3\sigma$ value.}
\end{table*}

\section{Results}
\label{Res}

\begin{figure}
   \centering
   \includegraphics[width=9cm]{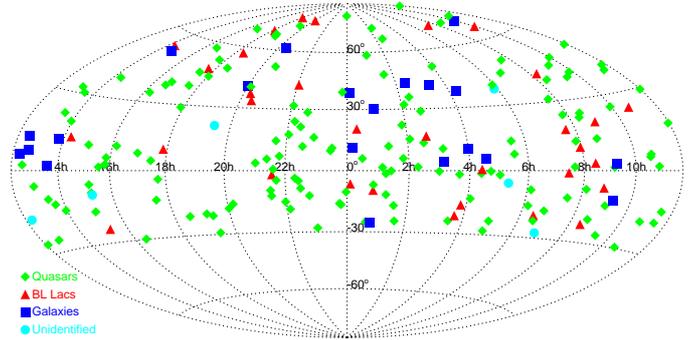}
   \caption{Sky distribution of sources in our sample in J2000.0 coordinates. }
   \label{skymap}
\end{figure}

Table~\ref{T2} shows the results of the new 86 and 229\,GHz observations presented here.
Together with the source name, we give the integration time, as well as the total flux density ($S_{\rm{86}}$ and $S_{\rm{229}}$ at 86 and 229\,GHz, respectively), the fractional linear polarization ($m_{\rm{L, 86}}$ and $m_{\rm{L, 229}}$), the linear polarization angle ($\chi_{\rm{86}}$ and $\chi_{\rm{229}}$), the fractional circular polarization at 86\,GHz ($m_{\rm{C, 86}}$), and the maximum Faraday rotation allowed from the $\chi_{\rm{86}}$ and $\chi_{\rm{229}}$ measurements.
For both 86 and 229\,GHz observations, $3\sigma$ upper limits in $S$, $m_{\rm{L}}$ and/or $m_{\rm{C}}$ were provided whenever the measurement did not exceed its corresponding $3\sigma$ value.
In Fig.~\ref{skymap} we also show the sky distribution of sources in our sample.

Note that whereas 86\,GHz linear polarization was detected from most sources in our sample (88\,\%), 86\,GHz circular polarization could be detected for only a small fraction of them (6\,\%).
At 229\,GHz, the total flux weakness of most sources in our sample, as well as the reduced sensitivity and sky transmission at this waveband, only allowed us to detect linear polarization from 13\,\% of our sample.
No circular polarization was detected at 229\,GHz.

In the following subsections we present the statistical analysis and discussion of the relevant aspects regarding these data.

\subsection{Total flux density}

\subsubsection{Total luminosity}
\begin{figure}
   \centering
   \includegraphics[width=9cm]{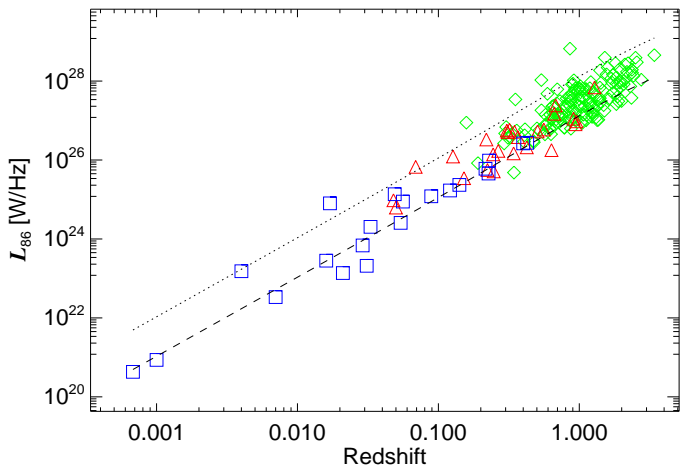}
   \caption{86\,GHz luminosity as a function of redshift. The dashed line indicates the luminosity for observer's frame flux density $S_{86}=0.5$\,Jy, whereas the dotted line is for $S_{86}=5$\,Jy. As for other figures hereafter, diamonds symbolize quasars, triangles denote BL~Lacs, and squares are radio galaxies.}
   \label{L86_z_QBG}
\end{figure}

\begin{figure}
   \centering
   \includegraphics[width=9cm]{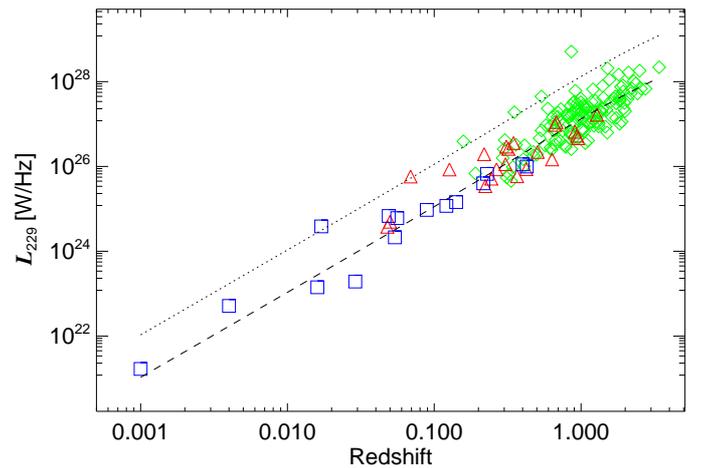}
   \caption{229\,GHz luminosity as a function of redshift. The dashed line indicates the luminosity for observer frame's flux density $S_{229}=0.5$\,Jy, whereas the dotted line is for $S_{229}=5$\,Jy.}
   \label{L229_z_QBG}
\end{figure}

Figures~\ref{L86_z_QBG} and \ref{L229_z_QBG} show the 86 and 229\,GHz luminosity ($L=4 \pi d_{L}^{2} S(1+z)^{-1}$, where $d_{L}$ is the luminosity distance\footnote{We use hereafter a $H_o=71$\,km\,s$^{-1}$\,Mpc$^{-1}$, $\Omega_{m}=0.27$ and $\Omega_{\Lambda}=0.73$ cosmology.}) as a function of redshift in our sample.
A significant fraction of sources (51\,\% [22\,\%]) show 86\,GHz flux densities bellow the 1\,Jy (0.5\,Jy) threshold, whereas only a small subset of 8 sources (4\,\%) shows $S_{86}>5$\,Jy.
At 229\,GHz, 80\,\% (48\,\%) of the sources show flux densities below 1\,Jy (0.5\,Jy), and only 6 of them (3\,\%) are brighter than 5\,Jy.
As expected, cosmologically distant quasars show the largest luminosities with median $\tilde{L}_{Q}^{86}=$2.7$\times10^{27}$\,W/Hz ($\tilde{L}_{Q}^{229}=$1.6$\times10^{27}$\,W/Hz), followed by BL~Lac objects with $\tilde{L}_{B}^{86}=$3.8$\times10^{26}$\,W/Hz ($\tilde{L}_{B}^{229}=$1.9$\times10^{26}$\,W/Hz), and radio galaxies $\tilde{L}_{G}^{86}=$8.0$\times10^{24}$\,W/Hz ($\tilde{L}_{G}^{229}=$6.8$\times10^{24}$\,W/Hz).
This is consistent with expectations from the AGN unification theory \citep{Urry:1995p6301}, where quasars and BL~Lacs are the relativistically Doppler boosted (i.e. beamed) versions of high power and low power radio galaxies (with their jets oriented closer to the plane of the sky), respectively. 

\subsubsection{Spectral index}
\label{alpha}

To study the spectral differences between quasars and BL~Lacs, we determined the 15 to 86\,GHz spectral index\footnote{We define the spectral index between two observing frequencies $\nu_{1}$ and $\nu_{2}$ as $\alpha_{\nu_{1}{\rm,}\nu_{2}}=log(S_{\nu_{1}}/S_{\nu_{2}})/log({\nu_{1}}/{\nu_{2}})$ .} ($\alpha_{15{\rm,}86}$) of quasars and BL Lacs. 
For that we used the 15 GHz total flux densities from integrated intensities in available MOJAVE VLBA images\footnote{\tt http://www.physics.purdue.edu/MOJAVE} with observations performed as contemporaneous as possible to our millimeter observations.

\begin{figure}
   \centering
   \includegraphics[width=9cm]{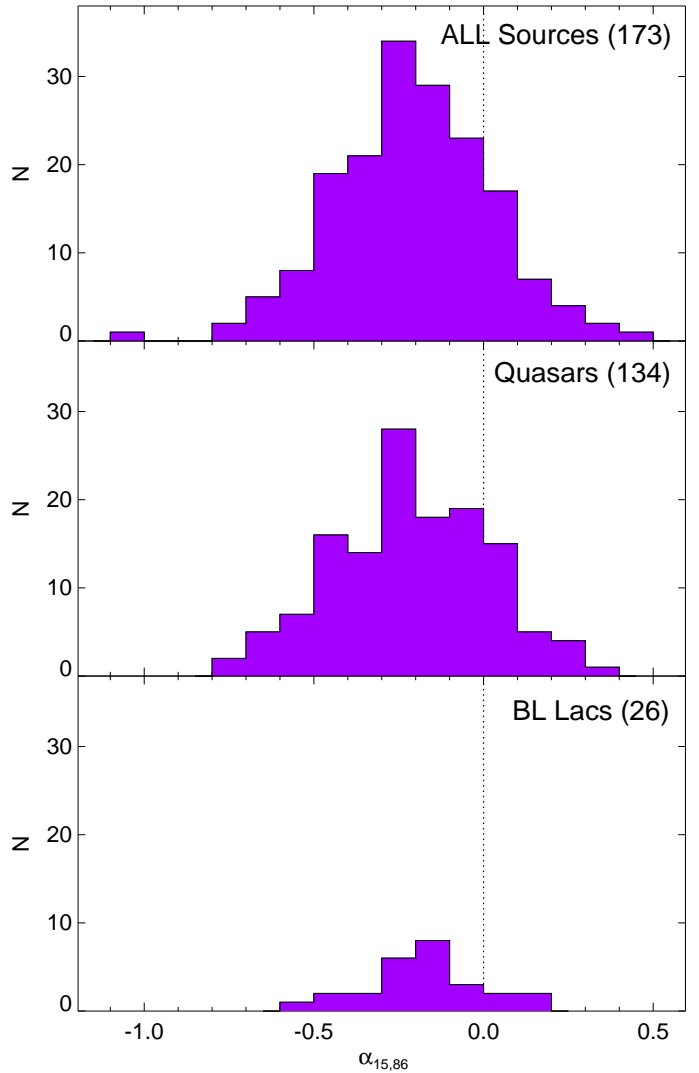}
   \caption{Distribution of 15\,GHz to 86\,GHz spectral indices ($\alpha_{15{\rm,}86}$) for all sources in both the MOJAVE and in our sample, and their corresponding quasar, and BL~Lac subsamples. The 15\,GHz total flux density was taken from integrated intensity of MOJAVE images. For each source, the 15\,GHz observation take at the closest date to our 86\,GHz measurement was selected. Numbers in parentheses denote sample sizes.}
   \label{spindx86_15}
\end{figure}

Similar to what it was found in \citet{Agudo:2010p12104}, Fig.~\ref{spindx86_15} shows that the $\alpha_{15{\rm,}86}$ spectral indices for both quasars and BL~Lacs are distributed towards flat and optically thin spectral indices (with $\alpha_{15{\rm,}86}$ median $\tilde\alpha_{15{\rm,}86}^{\rm{Q}}=-0.22$ for quasars and $\tilde\alpha_{15{\rm,}86}^{\rm{B}}=-0.12$ for BL~Lacs).
Only a small fraction of sources (19\,\% of quasars and 15\,\% of BL~Lacs) show $\alpha_{15{\rm,}86}>0$.
The smallest average spectral index of quasars with regard to that of BL~Lacs is consistent with the well known trend for BL~Lacs to distribute the peaks of their synchrotron spectral energy distributions towards higher frequencies than quasars \citep[e.g.,][]{Ackermann:2011p17255,Giommi:2013p17953}. 
We also attribute part of the differences in spectral index between quasars and BL~Lacs to the larger cosmological redshifts of quasars that shifts their spectra to lower frequencies in the observer's frame.

\begin{figure}
   \centering
   \includegraphics[width=9cm]{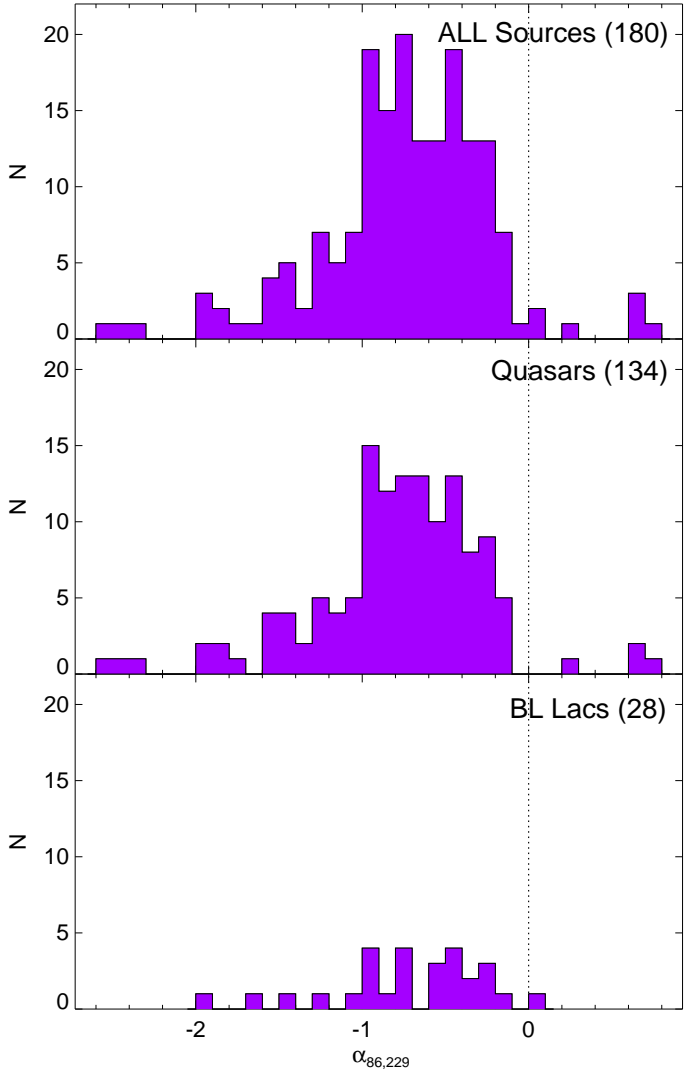}
   \caption{Distribution of 86\,GHz to 229\,GHz spectral indices ($\alpha_{86{\rm,}229}$) for all sources detected at both frequencies in our sample.}
   \label{spindx229_86}
\end{figure}

The 86 to 229 spectral index distribution ($\alpha_{86{\rm,}229}$, see Fig.~\ref{spindx229_86}) also shows a similar trend with BL~Lacs distributed towards slightly smaller spectral indexes with regard to quasars, although for the case of $\alpha_{86{\rm,}229}$, both the quasar and the BL~Lac samples distribute toward significantly smaller (more optically thin) spectral indexes (with $\tilde\alpha_{86{\rm,}229}^{\rm{Q}}=-0.75$ for quasars and $\tilde\alpha_{86{\rm,}229}^{\rm{B}}=-0.56$ for BL~Lacs).
These values are consistent with those for optically thin synchrotron radiation from AGN jets \citep{Rybicki:1979p6159}, which shows that blazars display in general optically thin radiation between 86 and 229\,GHz, and also guarantees that our polarimetric observations were not significantly affected by polarization angle rotation and depolarization owing to opacity effects.
There are however a few exceptions, in particular 21 our of 180, for which the spectral index is flat or optically thick (with $\alpha_{86{\rm,}229}\gtrsim0.25$), perhaps because of ongoing prominent flaring states \citep[e.g.][]{1985ApJ...298..114M}.

\subsection{Linear polarization}

\begin{figure}
   \centering
   \includegraphics[width=9cm]{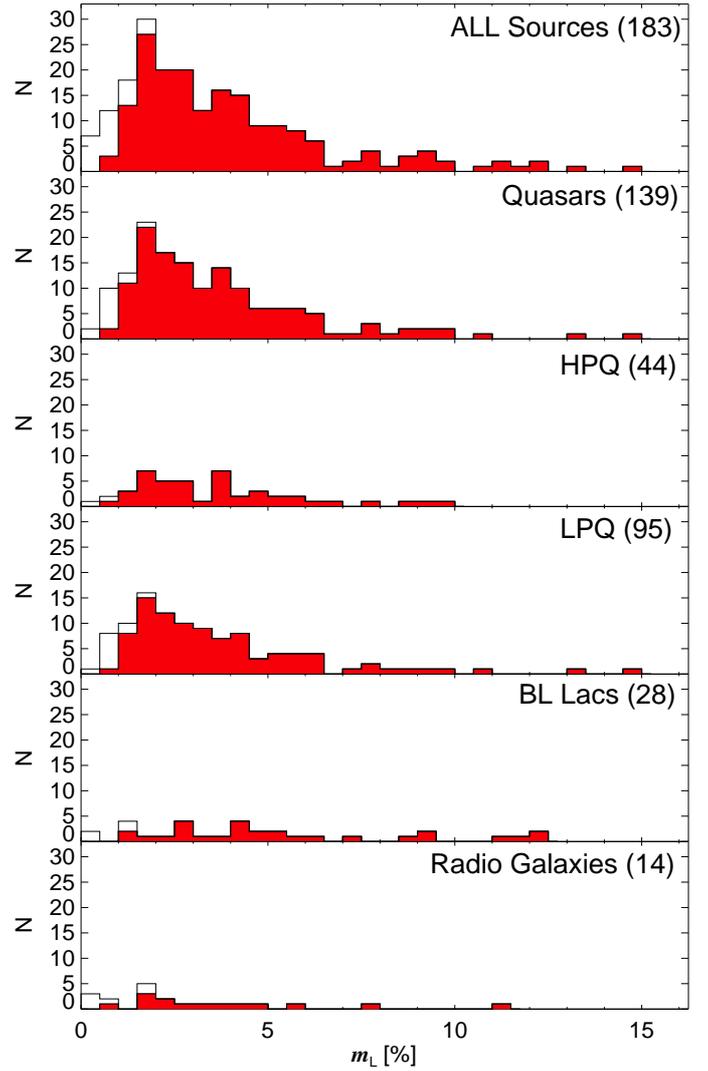}
   \caption{Distribution of 86\,GHz fractional linear polarization (red areas) for: all sources in the sample, quasars, high optical polarization quasars (HPQ), low optical polarization quasars (LPQ), BL~Lacs, and radio galaxies, from top to bottom. N is the number of sources in $0.5$\,\% wide bins. Non-filled areas also include non-detection upper limits. Note that since we chose $3\sigma$ upper limits with mean $\bar\sigma_{m_{\rm{L,86}}}\approx0.32$\,\%, hence there are only a few detections at $m_{\rm{L,86}}\lesssim0.96$\,\%. }
   \label{mALL86}
\end{figure}

\begin{figure}
   \centering
   \includegraphics[width=9cm]{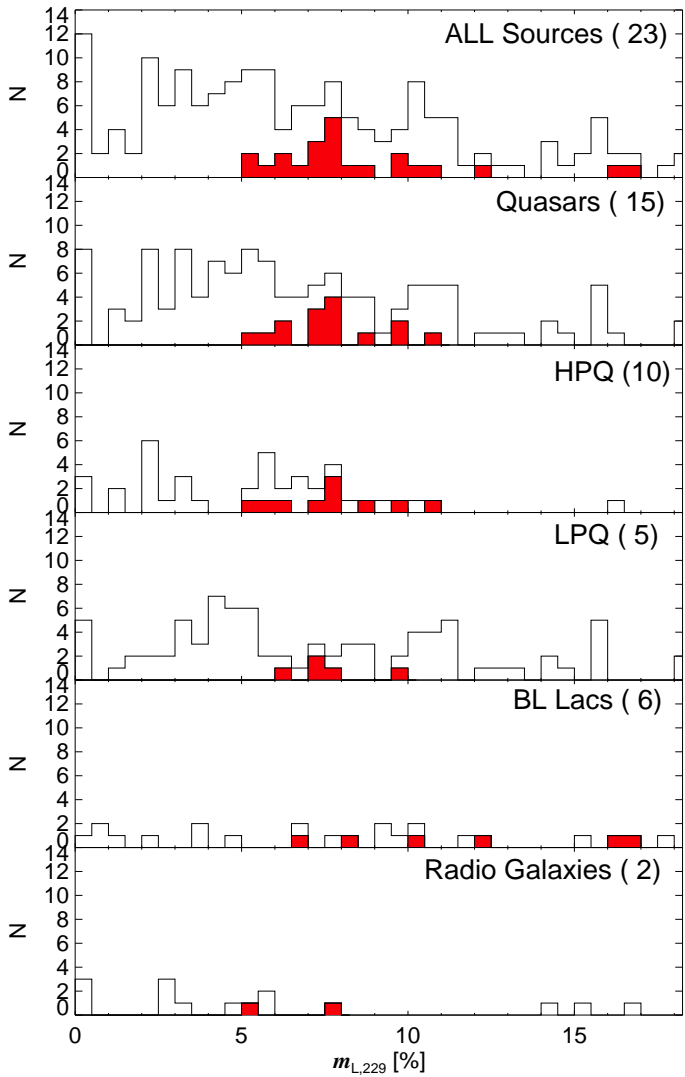}
   \caption{Same as Fig.~\ref{mALL86} but for 229\,GHz fractional linear polarization. Since we chose $3\sigma$ upper limits with mean $\bar\sigma_{m_{\rm{L,229}}}\approx2.0$\,\%, there are only a few detections at $m_{\rm{L,229}}\lesssim6.0$\,\% is available. }
   \label{mALL229}
\end{figure}

Fractional linear polarization at 86\,GHz ($m_{\rm{L, 86}}$) was detected at $\ge3\sigma$ level for 183 sources, a 88\,\% of the entire sample detected in total flux at 86\,GHz (Fig.~\ref{mALL86}).
At 229\,GHz, the reduced sensitivity of our observations as compared to those at 86\,GHz, only allowed to detect 23 sources, out of the 181 sources detected at 229\,GHz --a 13\,\% detection rate-- (Fig.~\ref{mALL229}).

We find the median values of $m_{\rm{L,86}}$ for BL~Lacs ($\tilde{m}_{\rm{L,86}}^{\rm{B}}=4.6$\,\%) to be appreciably larger than those for quasars ($\tilde{m}_{\rm{L,86}}^{\rm{Q}}=3.2$\,\%).
A similar result is found for the 229\,GHz linear polarization degree, where ($\tilde{m}_{\rm{L,229}}^{\rm{B}}=12.0$\,\%) is also much larger than for quasars ($\tilde{m}_{\rm{L,86}}^{\rm{Q}}=7.7$\,\%).
To make these comparisons, $3\sigma$ upper limits of  values of ${m}_{\rm{L}}$ were considered for non detections of linear polarization to avoid overestimation of the median of ${m}_{\rm{L}}$.
The difference of the BL~Lac and quasar distributions of linear polarization degree is confirmed by the Gehan's generalized Wilcoxon (GGW) test\footnote{The Gehan's generalized Wilcoxon test considers both detections and upper limits. To perform our tests, we used the ASURV~1.2 survival analysis package \citep[see ][ and references therein]{Lavalley:1992p3731}.} at a 97.1\,\% and 99.9\,\% confidence level for the 86 and 229\,GHz distributions, respectively.
This result (i.e., quasars significantly less polarized than BL~Lacs at millimeter wavelengths) was already pointed out at 86\,GHz in \citet{Agudo:2010p12104} and is also proved here for the first time at 229\,GHz.

The differences of linear polarization degree of quasars and BL~Lacs cannot be explained by differences in opacity in these two classes of sources since we have already shown that both of them are, in general, optically thin between 86 and 229\,GHz for most sources in their corresponding samples.
As in \citet{Agudo:2010p12104}, we consider a more plausible explanation the fact that quasars have, in general, smaller viewing angles than BL~Lacs  \citep{Hovatta:2009p3860,Pushkarev:2009p9412}, which produces stronger depolarization in quasars if the magnetic fields in the jets of blazars are not homogeneously distributed along the jet axis or the jets themselves are not axisymmetric.

In \citet{Agudo:2010p12104} we pointed out an apparent dichotomy in the 86\,GHz distribution of linear polarization degree of our entire source sample, which showed a peak at $m_{\rm{L,86}}\approx2.5$\,\%, and a second one at $m_{\rm{L,86}}\approx4$\,\%.
This behavior was also found at lower polarization degrees by \cite{Lister:2005p261} for the cores of quasars and the integrated polarization degree of EGRET-detected blazars.
The data from our new survey further confirms the dichotomy found in \citet{Agudo:2010p12104}.
The  $m_{\rm{L,86}}$ distribution of the entire source sample shown in Fig.~\ref{mALL86} also shows a possible double peaked-like shape with the first peak in the range $m_{\rm{L,86}}\in[1.5,2.5]$\,\%, and the second one at $m_{\rm{L,86}}\approx4$\,\%.
The lack of enough polarization sensitivity of our 229\,GHz observations with regard to those at 86\,GHz prevents us to study this dichotomy at 229\,GHz, where linear polarization below the  5\,\% level is not detected (Fig.~\ref{mALL229}).

The similarity of the 86\,GHz  $m_{\rm{L,86}}$ distributions of the entire source sample and that of quasars (which however only has a confidence of 78.9\,\% according to our GGW test) might indicate that this dichotomy is produced by quasars, but not by BL~Lacs (which $m_{\rm{L,86}}$ distribution is significantly different than those of quasars, see above).
Here we test the hypothesis that the bimodal $m_{\rm{L,86}}$ is produced by high optical polarization quasars (HPQ, for the high polarization peak in the $m_{\rm{L,86}}$ distribution of quasars) as listed in  \citet{VeronCetty:2006p4900}  and low optical polarization quasars (LPQ, for the low polarization peak).
This, however, does not seem to be a reliable explanation, given that the HPQ $m_{\rm{L,86}}$ distribution is not clearly accumulated towards large linear polarization degrees, and we do not have the means to statistically demonstrate that the HPQ and LPQ  $m_{\rm{L}}$ distributions are significantly different (the GGW test gives only a 52,6\,\% confidence for that).
Therefore, although we have been able to confirm the dichotomy in the $m_{\rm{L,86}}$ distribution of our entire source sample (quasars in particular), our new data remain inconclusive about the possible physical meaning of this apparent dichotomy.

\subsubsection{Fractional Linear Polarization Ratio Along the Radio and Millimeter Spectrum}

The linear polarization detection rate in our survey is $\sim88$\,\% and $\sim13$\,\% at 86 and 229\,GHz, respectively (see above).
Since we have chosen $3\sigma$ criteria for detection of linear polarization at both 86 and 229\,GHz --with median $3\tilde{\sigma}_{m_{\rm{L,86}}}\sim1.0$\,\%, and $3\tilde{\sigma}_{m_{\rm{L,229}}}\sim6.0$\,\%, respectively over the entire sample of detected sources at each observing frequency--, $\sim88$\,\% of our sources detected in total flux at 86\,GHz, show $m_{\rm{L,86}}\gtrsim1.0$\,\%, and $\sim13$\,\% of those detected in total flux at 229\,GHz, show $m_{\rm{L,229}}\gtrsim6.0$\,\%.

In contrast, 71\,\% of sources both in the MOJAVE and in our sample of 86\,GHz detections show $m_{\rm{L,86}}\gtrsim1.0$\,\%, whereas only 8\,\% of MOJAVE sources with linear polarization detected at 229\,GHz display $m_{\rm{L,229}}\gtrsim6.0$\,\%.
This points out a progressive increasing dependence of the fractional linear polarization of blazars with observing frequency \citep[see also][]{Agudo:2006p203,2007AJ.134.799J,Agudo:2010p12104} can also be discerned from Figs.~\ref{mALL86} and \ref{mALL229}.

\begin{figure}
   \centering
   \includegraphics[width=9cm,clip]{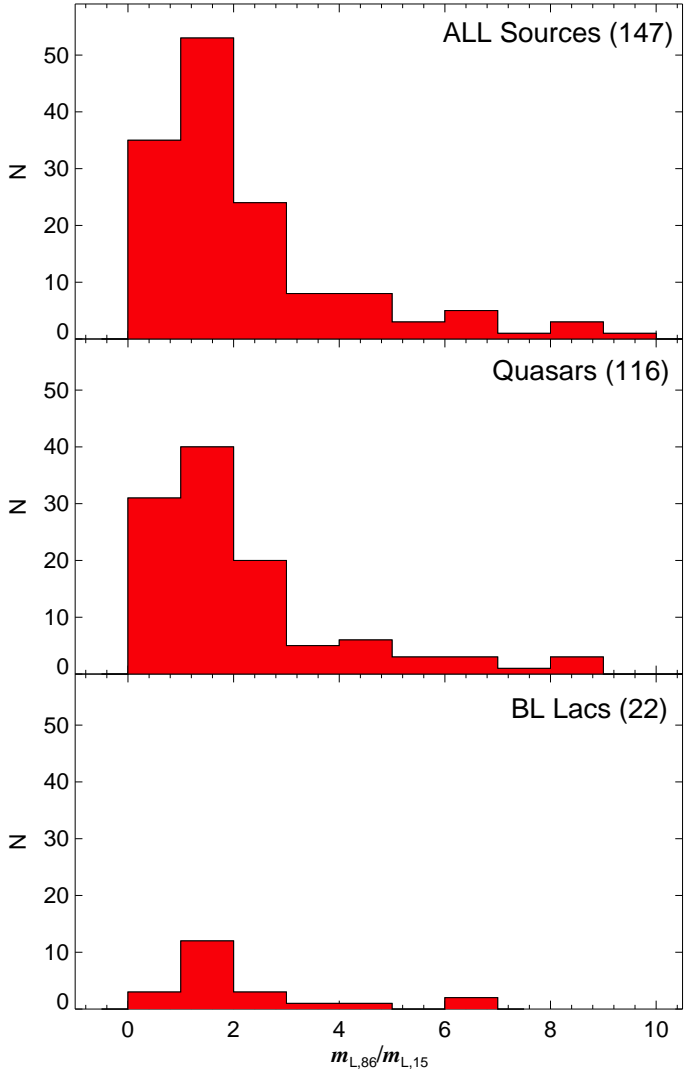}
   \caption{Distribution of 86\,GHz to 15\,GHz fractional linear-polarization ratio for sources with detected linear polarization both in our survey and in MOJAVE. Six sources in the range $10<m_{\rm{L,86}}$/$m_{\rm{L,15}}<19$ are not shown. The 15\,GHz linear polarization fraction was computed from measurements of integrated total flux density and linearly polarized flux density from 15\,GHz VLBI images taken by the MOJAVE team on dates contemporaneous to our 86\,GHz observations.}
   \label{ml86_ml15}
\end{figure}

\begin{figure}
   \centering
   \includegraphics[width=9cm,clip]{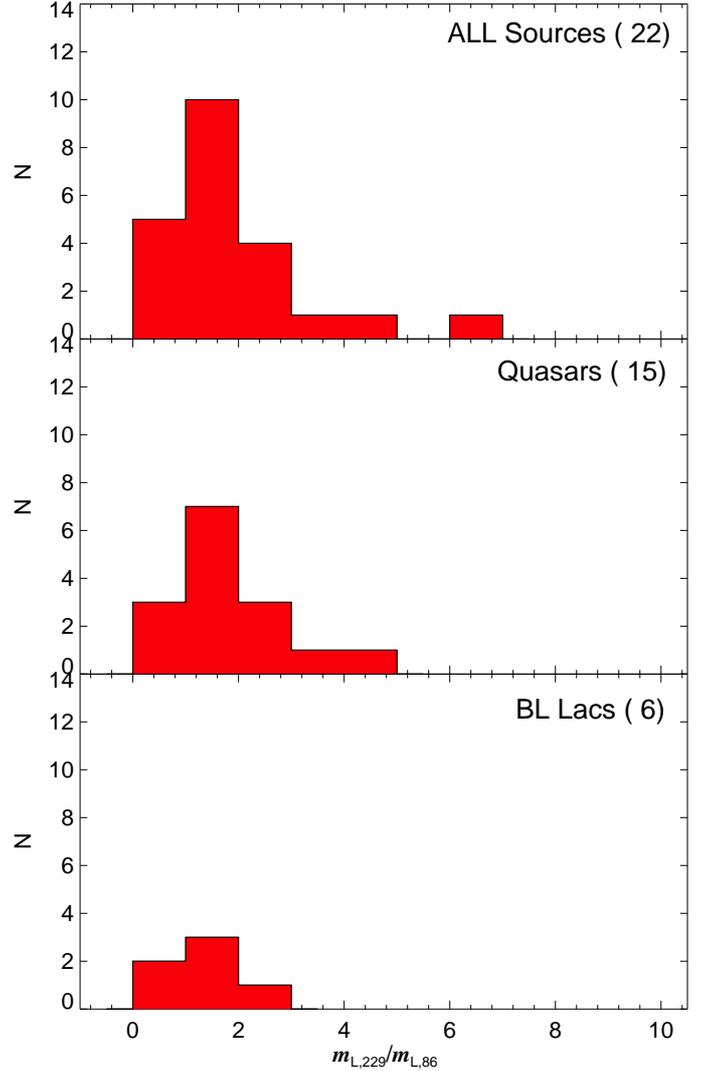}
   \caption{Distribution of 229\,GHz to 86\,GHz fractional linear-polarization ratio for sources with detected linear polarization at both frequencies (i.e., $m_{\rm{L,229}}$/$m_{\rm{L,86}}$).}
   \label{ml229_ml86}
\end{figure}

By using the data from our first 3.5 AGN Polarimetric Survey in \citet{Agudo:2010p12104} we demonstrated that the 86\,GHz linear polarization degree of blazars was, in general, around twice that at 15\,GHz.
Figure~\ref{ml86_ml15}, where we plot the distributions of $m_{\rm{L,86}}/m_{\rm{L,15}}$ for those sources detected at 86\,GHz in our survey and also with available contemporaneous 15\,GHz MOJAVE measurements of the linear polarization degree, also show the same result. 
The entire source sample, quasars and BL~Lacs show significantly larger fractional linear polarization at 86\,GHz than at 15\,GHz by a mean factor $\approx2$ (Fig.~\ref{ml86_ml15}).
Essentially the same result is found for the distributions of $m_{\rm{L,229}}/m_{\rm{L,86}}$, when the linear polarization of sources detected at 229\,GHz in our survey is compared with their corresponding -- simultaneously measured-- 86\,GHz linear polarization degree (Fig.~\ref{ml229_ml86}).
The median values of the $m_{\rm{L,86}}/m_{\rm{L,15}}$ and $m_{\rm{L,229}}/m_{\rm{L,86}}$ distributions show, however, slightly smaller values, being 1.6, 1.6, and 1.5 for the entire sample, quasars and BL~Lacs, respectively for the case of $m_{\rm{L,86}}/m_{\rm{L,15}}$, and 1.7, 1.7, and 1.5 for $m_{\rm{L,229}}/m_{\rm{L,86}}$.

Although there are some apparent differences on the different distributions shown in Figs.~\ref{ml86_ml15} and \ref{ml229_ml86}, our Kolmogorov--Smirnov (K-S) tests do not give sufficiently high confidence to conclude that quasars and BL~Lacs distributions of $m_{\rm{L,86}}/m_{\rm{L,15}}$ and $m_{\rm{L,229}}/m_{\rm{L,86}}$ are selected from different parent distributions (confidence level only 40.9\,\%, and 88.3\,\%, respectively).
Note that there is also a prominent 18\,\% (9\,\%)  fraction of sources with larger $m_{\rm{L,86}}/m_{\rm{L,15}}$ ($m_{\rm{L,229}}/m_{\rm{L,86}}$) ratio than $4$.
Such tail of large high-frequency linear-polarization excess seems to be present in both quasars and BL~Lacs, although BL~Lacs do not show it so clearly in $m_{\rm{L,229}}/m_{\rm{L,86}}$, perhaps because of the lower number of sources in that subsample.

When we reported for the first time the factor of $\approx2$ of $m_{\rm{L,86}}$ with regard to $m_{\rm{L,15}}$ \citep{Agudo:2010p12104}, we proposed two complementary explanations for this effect: \emph{a)} that the 86\,GHz emission in blazars comes from a region with better ordered magnetic field than the 15\,GHz one, and \emph{b)} that the 15\,GHz emission from blazars is affected by considerably greater Faraday depolarization relative to the 86\,GHz emission.
Now that we have also studied the properties of the $m_{\rm{L,229}}/m_{\rm{L,86}}$ distributions in this work, we can safely assume that explanation \emph{b)} is not reliable, at least for the comparison of the 86 and the 229 GHz emission, that is very unlikely to show appreciable Faraday rotations for a significant number of sources in our samples (see Sections~\ref{Intr} and \ref{Far}).
Option \emph{a)}, and the existing evidence that higher millimeter emission from blazars comes from inner regions upstream in their jets \citep[e.g.,][]{2007AJ.134.799J}, imply that the magnetic field is progressively better ordered in blazar jet regions located progressively upstream in the jet.

\subsubsection{Total Luminosity vs. Linear Polarization}
\label{S_vs_pL}

\begin{figure}
   \centering
   \includegraphics[width=9cm]{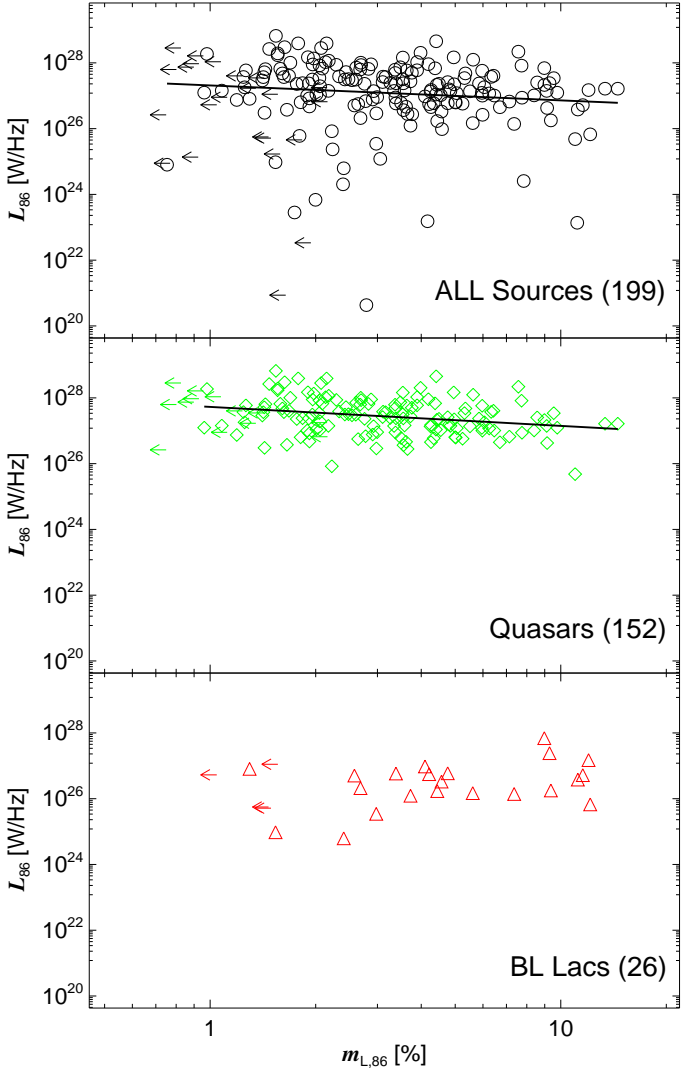}
   \caption{86\,GHz luminosity versus fractional linear polarization at 86\,GHz for sources with known redshift for the entire source sample, quasars, and BL~Lacs (from top to bottom). Arrows symbolize $m_{{\rm{L}}}$ upper limits. The continuous lines symbolize the result of linear regressions. Numbers in parentheses denote sample sizes.}
   \label{L86_ml86}
\end{figure}

In \citet{Agudo:2010p12104}, we reported for the first time a significant anti--correlation between the 86\,GHz luminosity ($L_{86}$) and $m_{{\rm{L,86 }}}$ for blazars in our previous 3.5\,mm AGN Polarization Survey.
The data from our new survey also reproduces such anti--correlation.
Fig.~\ref{L86_ml86} shows $L_{86}$ versus $m_{{\rm{L,86 }}}$ for sources with known redshift in the entire source sample and for quasars and BL~Lacs.
Our Spearman's $\rho$ test for correlation gives $\rho=-0.22$ with 99.7\,\% confidence for the entire source sample and $\rho=-0.27$ with 99.9\,\% confidence for quasars.
Perhaps because of the soft dependence of $L_{86}$ versus $m_{{\rm{L,86 }}}$, and also because of the small number of measurements for BL~Lacs, this anti--correlation cannot be confirmed for BL~Lacs, for which $\rho=0.25$ with 74.5\,\% confidence.
A similar analysis performed through the Generalized Spearman's $\rho$ test -- using the ASURV~1.2 package by \citet{Lavalley:1992p3731}, thus taking into account both detections and upper limits -- also points out statistically significant anti--correlation for quasars, with $\rho=-0.27$ at 99.9\,\% confidence.
The Generalized Spearman's $\rho$ test only gives 91.2\,\% confidence though for $\rho=-0.12$ for the entire source sample when upper limits are accounted for, and even less ($\rho=0.24$ at 76.9\,\% confidence) for BL~Lacs.
At 229\,GHz, the small number of linear polarization detections do not allow for reliable correlation studies of $L_{229}$ versus $m_{{\rm{L,229}}}$.

\begin{figure}
   \centering
   \includegraphics[width=9cm]{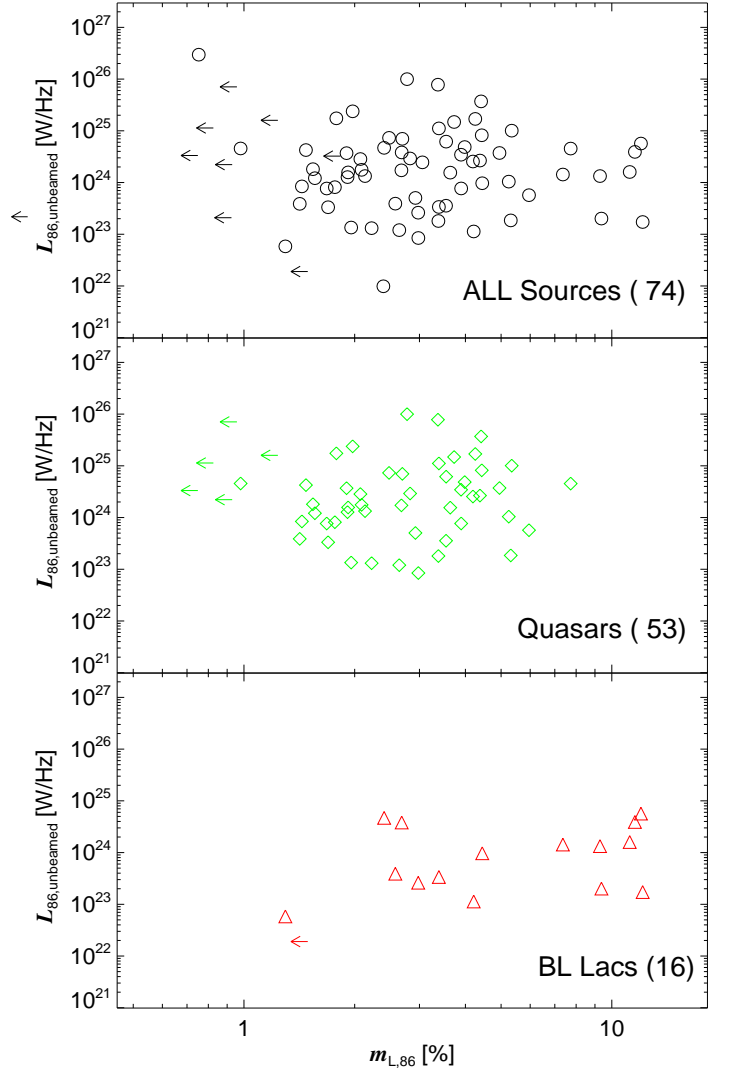}
   \caption{Same as Fig.~\ref{L86_ml86} but for 86\,GHz unbeamed luminosities computed from Doppler factors given in \cite{Hovatta:2009p3860}.}
   \label{L86unb_ml86}
\end{figure}

The confirmation of this result implies that the magnetic field order in jets of blazars increases with decreasing millimeter luminosity.
In \citet{Agudo:2010p12104} we tested the hypothesis that the  $L$ versus $m_{{\rm{L}}}$ anti--correlation would be only produced by orientation and relativistic effects, i.e. sources whose jets are better oriented to the line of sight are expected to display larger luminosities (because of their larger Doppler boosting) and also lower linear polarization degrees (because of cancellation of orthogonal polarization components along the line of sight).
This was tested by computing the unbeamed luminosities ($L_{\rm{86,unbeamed}}$) of sources with known Doppler factors from \cite{Hovatta:2009p3860}, but this hypothesis was ruled out because significant correlation was still found for a single subsample of sources.
We have repeated the same test with our updated data set, and we have obtained no correlation between the $L_{\rm{unbeamed}}$ and $m_{{\rm{L}}}$ (Fig.~\ref{L86unb_ml86}) neither for the entire source sample, nor for quasars or BL~Lacs. 
This opens again the possibility to use orientation and relativistic effects to explain the reported $L_{\rm{beamed}}$ versus $m_{{\rm{L}}}$ anti--correlation, although other alternative explanations \citep[e.g.][]{Agudo:2010p12104}, cannot be ruled out.

\subsubsection{Linear polarization angle vs. jet position angle}
\label{Misal}

\begin{figure}
   \centering
   \includegraphics[width=8.5cm]{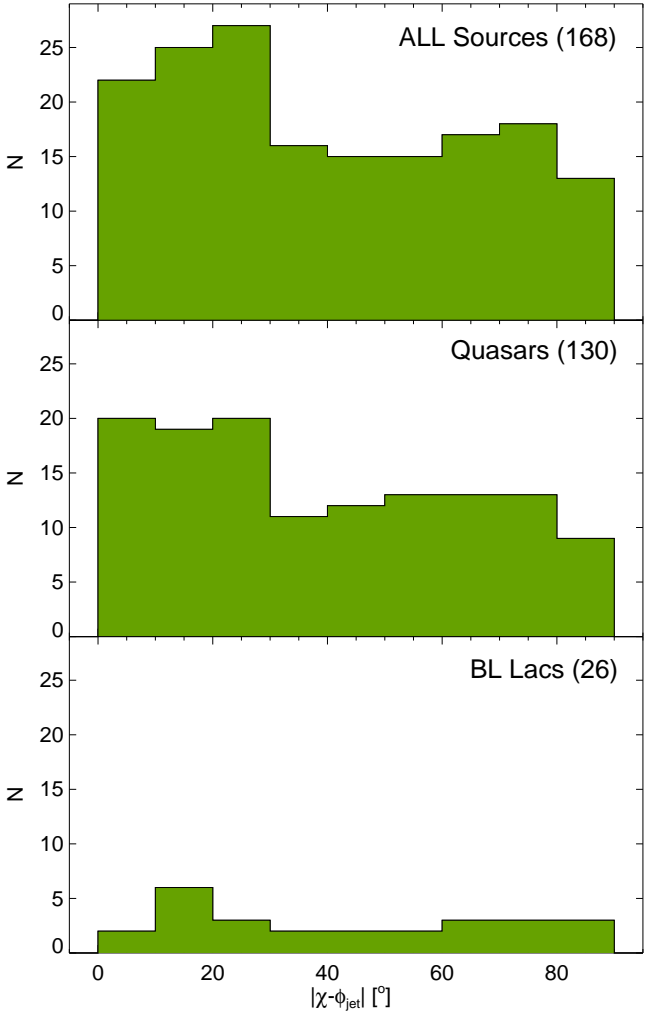}
   \caption{Distribution of misalignment between $\chi_{86}$ and $\phi_{\rm{jet}}$. We present, from top to bottom, the entire source sample and the quasar and BL~Lac subsamples.}
   \label{misal_86}
\end{figure}

Figure~\ref{misal_86} shows the distribution misalignment of jet position angle ($\phi_{\rm{jet}}$, see Table~\ref{T1}) with polarization (electric vector) position angle at 86\,GHz ($\chi_{86}$, see Table~\ref{T2}), i.e., $|\chi_{86}-\phi_{\rm{jet}}|$, for the entire source sample, quasars, and BL~Lacs.
To give an estimate of $\phi_{\rm{jet}}$ as reliable as possible to compare with our millimeter polarimetric measurements, we first searched in the 86\,GHz VLBI Survey by \citet{Lee:2008p301}, followed by the 15\,GHz data from the MOJAVE survey \citep[][preferentially]{Lister:2005p261}.
A more exhaustive search from references 1 to 10 on Table~\ref{T1} was done for every source not found in these two surveys. 
If a source was found in several references, we adopted the higher frequency $\phi_{\rm{jet}}$ measurement.

\begin{figure}
   \centering
   \includegraphics[width=8.5cm]{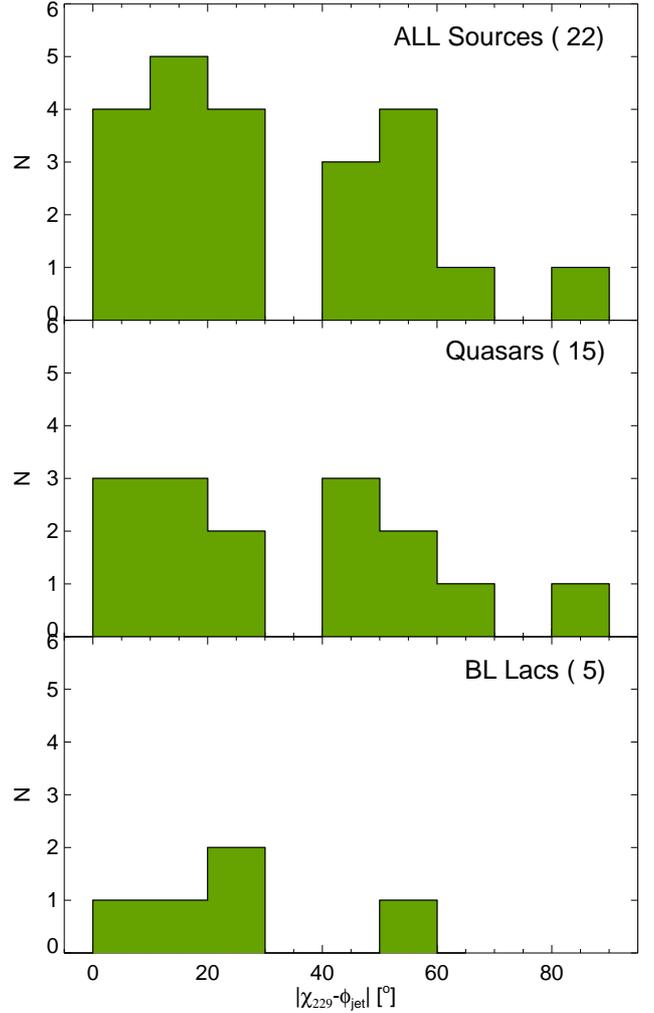}
   \caption{Sama as Fig~\ref{misal_86} but for 230\,GHz linear polarization angle ($\chi_{230}$) measurements.}
   \label{misal_230}
\end{figure}

Figure~\ref{misal_86} shows a very weak --almost inexistent-- trend for sources in our entire sample and quasars to distribute $\chi_{86}$ nearly parallel to $\phi_{\rm{jet}}$, with $0^{\circ}\le|\chi_{86}-\phi_{\rm{jet}}|\lesssim30^{\circ}$. 
A similar result was also shown in \citet{Agudo:2010p12104}.
BL~Lacs seem to show a different $|\chi_{86}-\phi_{\rm{jet}}|$ distribution with an additional (but also very weak)excess of sources with $\chi_{86}$ almost parallel to $\phi_{\rm{jet}}$.
However, our Kolmogorov-Smirnov (K-S) tests give a too weak confidence on the hypothesis that our entire source sample and quasars come from a different parent distribution than BL~Lacs (12\,\%, and 23\,\%, respectively), and therefore it is not possible to confirm this difference.
Our 229\,GHz linear polarization angle measurements give similar results (Fig.~\ref{misal_230}).
We report an also very weak trend to distribute the 229\,GHz polarization angle nearly parallel to the jet position angle for the entire source sample and quasars.
The K-S test on their $|\chi_{229}-\phi_{\rm{jet}}|$ distributions give a 99.5\,\% confidence to come from the same parent distribution.
For BL~Lacs $|\chi_{229}-\phi_{\rm{jet}}|$ might be distributed distributed in all possible misalignment angles, although the small number of sources in this case (Fig.~\ref{misal_230}) do not allow us to make any reliable statement or statistical tests.

Even for the case of the entire source sample considered for the $|\chi_{86}-\phi_{\rm{jet}}|$ histogram, the excess of sources accumulated towards $0^{\circ}\le|\chi_{86}-\phi_{\rm{jet}}|\lesssim30^{\circ}$ only represents a 17\,\% of the total number of sources in our entire sample, which gives an idea of the small relevance of this excess.
Therefore, we fully confirm that there is no clear trend for the millimeter linear polarization angle to be aligned either parallel of perpendicular to $\phi_{\rm{jet}}$ for all source sub-samples considered here.
This result contradicts theoretical expectations for axisymmetric jets, which predict that the polarization angle should be observed either parallel or perpendicular to the jet axis \citep[e.g,][]{Lyutikov:2005p321,Cawthorne:2006p409}.
Also, several previous observational attempts to probe this bi-modality through observations at centimeter wavelengths \citep[e.g.][]{2000MNRAS.319.1109G,2003ApJ...589..733P,Lister:2005p261} do not show agreement among each other \citep[see][ for a summary on their differences]{Agudo:2010p12104}.

An explanation why we do not detect a clear trend of either quasars or BL~Lacs to distribute their short millimeter polarization degree either parallel or perpendicular to the jet might be:
a) a larger $\chi$ variability amplitude and/or time scale at millimeter wavelengths with regard to those at longer centimeter wavelengths,
b) different physical properties of the region where the bulk of the short millimeter emission is radiated (with regard to those at longer centimeter wavelengths),
c) and significant departures from axisymmetric jet geometries and dynamics on the short millimeter emission regions that should hence show different expected integrated polarization angles than those for axisymmetric jets.

\subsection{Faraday rotation between 86 and 229\,GHz}
\label{Far}

\begin{figure}
   \centering
   \includegraphics[width=8.5cm]{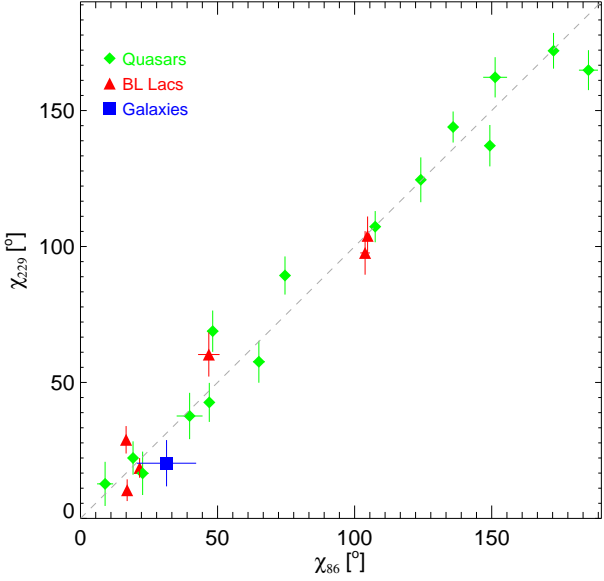}
   \caption{86\,GHz linear polarization angle ($\chi_{86}$) as compared to the simultaneously measured 229\,GHz linear polarization angle ($\chi_{229}$) for the 22 sources with detected polarization at both 86 and 229\,GHz. The dashed line represents the $\chi_{86}=\chi_{229}$ line. }
   \label{chi86_229}
\end{figure}

In Fig.~\ref{chi86_229} we compare the linear polarization angle measured at 86 and 229\,GHz for the 22 sources detected in polarization at both frequencies.
The Figure clearly shows that there is a general good match between  $\chi_{86}$ and $\chi_{229}$ within the errors.
Although there are some deviations from the $\chi_{86} = \chi_{229}$ line in Fig.~\ref{chi86_229}, none of the points are away from this line at more than $3\sigma$ with regard to the $\chi_{229}$ measurement.
The relatively large $\chi_{229}$ uncertainties -which are typically several times larger than those at 86\,GHz- do not allow us to provide a $>3\sigma$ measurement of Faraday rotation measure (RM)\footnote{$\rm{RM}=(\chi_{\rm{obs}}(\lambda)-\chi_{\rm{int}})/\lambda^{2}$, with $\chi_{\rm{int}}$ being the intrinsic polarization angle, and $\chi_{\rm{obs}}(\lambda)$ the observed polarization angle at the observing wavelength $\lambda$.} for any of the sources with both 86 and 229\,GHz polarization angle measurements.
Instead, only a $3\sigma\,\rm{RM}$ value --typically of the order of several times $10^4\,\rm{rad\,m}^2$-- is given in Table~\ref{T2} for the 22 sources mentioned above.
This result is in line with previous expectations for rotation measures not much larger than $\rm{RM}\sim10^4$\,rad\,m$^{-2}$, although with the available data we cannot rule out such large values. 
Large rotation measures have already been detected in some sources through ultra-high-resolution and high-precision  polarimetric-VLBI observations \citep[e.g.,][]{Attridge:2005p220,Gomez:2008p30,Gomez:2011p16108,Agudo:2012p17626}.

\section{Circular polarization}
\label{mC}

\begin{figure}
   \centering
   \includegraphics[width=8.5cm]{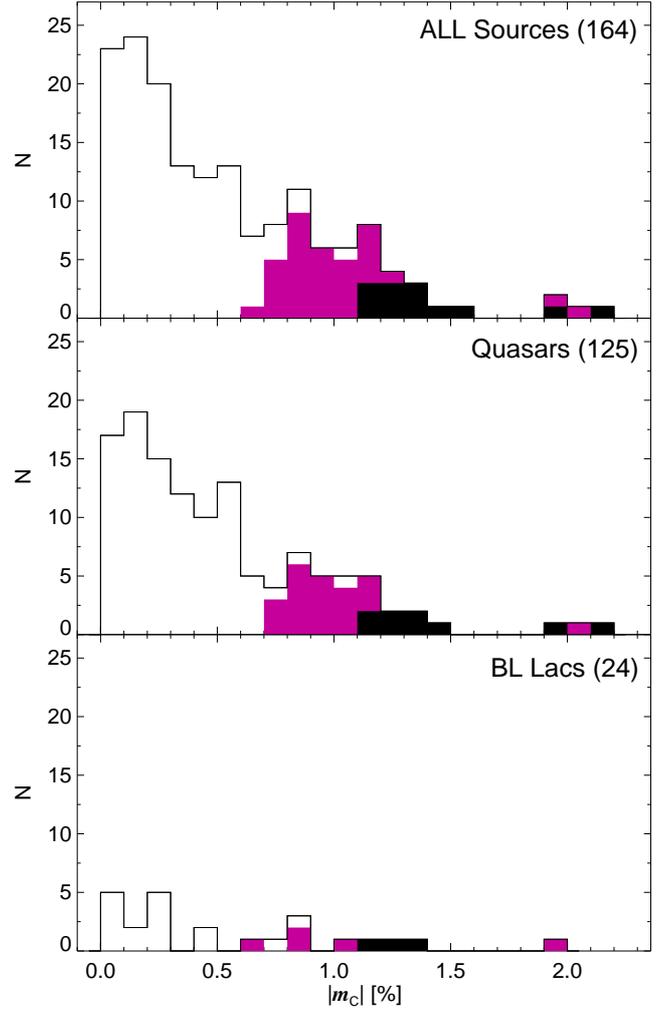}
   \caption{Distribution of the absolute value of 86\,GHz circular polarization for the entire source sample, quasars and BL~Lacs. Black areas correspond to $m_{\rm{C,86}}$ detections at $\ge3\sigma$. Violet shaded areas indicate observing results with $\ge2\sigma$, whereas unshaded areas symbolize all $m_{\rm{C}}$ measurements, regardless of their significance.}
   \label{mcALL}
\end{figure}

In Fig.~\ref{mcALL} we present the distribution of the absolute value of 86\,GHz circular polarization $|m_{\rm{C,86}}|$ for the samples considered here.
We show the distributions of $|m_{\rm{C,86}}|$ detections at $\ge3\sigma$ level,  $|m_{\rm{C,86}}|$ measurements at $\ge2\sigma$ level, and all $|m_{\rm{C,86}}|$ measurements (regardless of their significance).

As expected from the previously known low level of circular polarization degree of blazars \citep[typically $\lesssim0.5$\,\% at 2\,cm][]{Homan:2006p238}, we only detect $m_{\rm{C,86}}$ for a small fraction of sources (6\,\%).
This small detection rate is consistent with the $\sim15$\,\% one reported by \citet{Homan:2006p238} through 15\,GHz VLBA observations, and our own results for our previous survey  \citep{Agudo:2010p12104}, where we only detected circular polarization at 86\,GHz for 8 sources (6\,\% of the entire sample).
No circular polarization was detected from our 229\,GHz observations.

Figure~\ref{mcALL} shows that most of the 13 circular polarization detections correspond to values in the range $0.9\,\%\lesssim |m_{\rm{C,86}}|\lesssim1.6$\,\% (for sources 0229$+$131, 0430$+$052, 0451$-$282, 0716$+$714, 0917$+$449, 0954$+$658, 1324$+$224, 1328$+$307, 1642$+$690, 2131$-$021, and 2342$-$161), although there are two detections at $|m_{\rm{C,86}}|\approx2$\,\% (for 0923$+$392 and 1124$-$186).
These correspond to significantly larger $|m_{\rm{C,86}}|$ values than those observed on our first survey \citep{Agudo:2010p12104}, where we only detected 86\,GHz circular polarization in the range $0.3\,\%\lesssim|m_{\rm{C,86}}|\lesssim0.7$\,\%.
Apart from variability, we do not have an explanation why no large circular polarization values ($|m_{\rm{C,86}}|\gtrsim0.7$\,\%) was detected in our previous 86\,GHz survey, despite 8 sources detected in circular polarization in our new survey were also present in the sample of our previous survey.

Also, among all sources for which we detect 86\,GHz circular polarization in our new survey, only one (1124$-$186 with $m_{\rm{C,86}}=-1.98\pm0.35$\,\%) was also detected in the previous survey (with $m_{\rm{C,86}}=0.58\pm0.19$\,\%).
We also invoke circular polarization variability to explain this $m_{\rm{C,86}}$ sign reversal, as well as the large difference in $|m_{\rm{C,86}}|$ in five years between the observations of the two surveys.
This is consistent with circular polarization variability levels previously reported for radio loud AGN \citep[e.g.,][]{Aller:2003p4756}.

Moreover, among all sources with detected $m_{\rm{C,86}}$ in our new survey, only 0716$+$714 (with  $m_{\rm{C,86}}=-1.25\pm0.35$\,\%) was also detected by the MOJAVE team with (with  $m_{\rm{C,15}}=+0.37\pm0.11$\,\%), which is also consistent with models that can reproduce considerable  $m_{\rm{C}}$ differences at different observing frequencies \citep[e.g.,][]{Homan:2009p6162}.

\section{Total flux and linear polarization variability}
\label{Var}

\begin{figure}
   \centering
   \includegraphics[width=8.5cm]{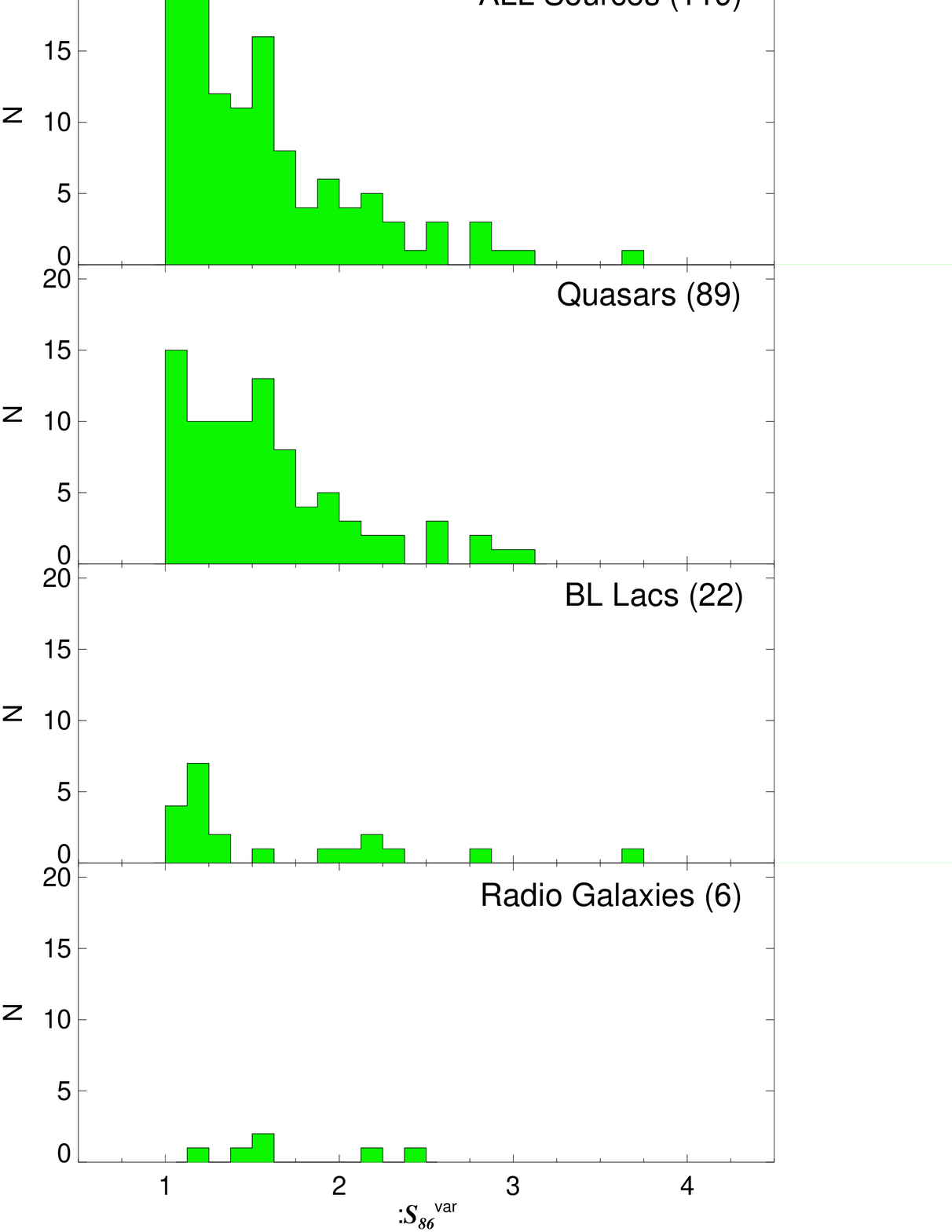}
   \caption{Distribution of 86\,GHz total flux variability ratio ($:S_{86}^{\rm{var}}$, see the text) for the entire source sample, quasars, BL~Lacs and radio galaxies. One source (0422+004) with $:S_{86}^{\rm{var}}=6.3$ is out of scale.}
   \label{Svar}
\end{figure}

In Fig.~\ref{Svar} we compare the 86\,GHz total flux density from the first 3.5\,mm AGN Polarimetric Survey in \citet{Agudo:2010p12104} and the measurements presented in this paper through the 86\,GHz total flux variability ratio ($:\,S_{86}^{\rm{var}}$)\footnote{$:\,S_{86}^{\rm{var}}=\frac{max(S_{86}^{\rm{Agudo\,et\,al.\,(2010)}},S_{86}^{\rm{this\,paper}})}{min(S_{86}^{\rm{Agudo\,et\,al.\,(2010)}},S_{86}^{\rm{this\,paper}}})$}.
The figure, clearly shows a large level of variability by a median factor of $\sim1.5$ for the entire source sample in time scales $\lesssim5$ years, and with a $19$\,\% of sources displaying $S_{86}$ variations by a factor $>2$.
No significant difference in the $:S_{86}^{\rm{var}}$ distribution is found between any of the different samples considered in Fig.~\ref{Svar}.

\begin{figure}
   \centering
   \includegraphics[width=8.5cm]{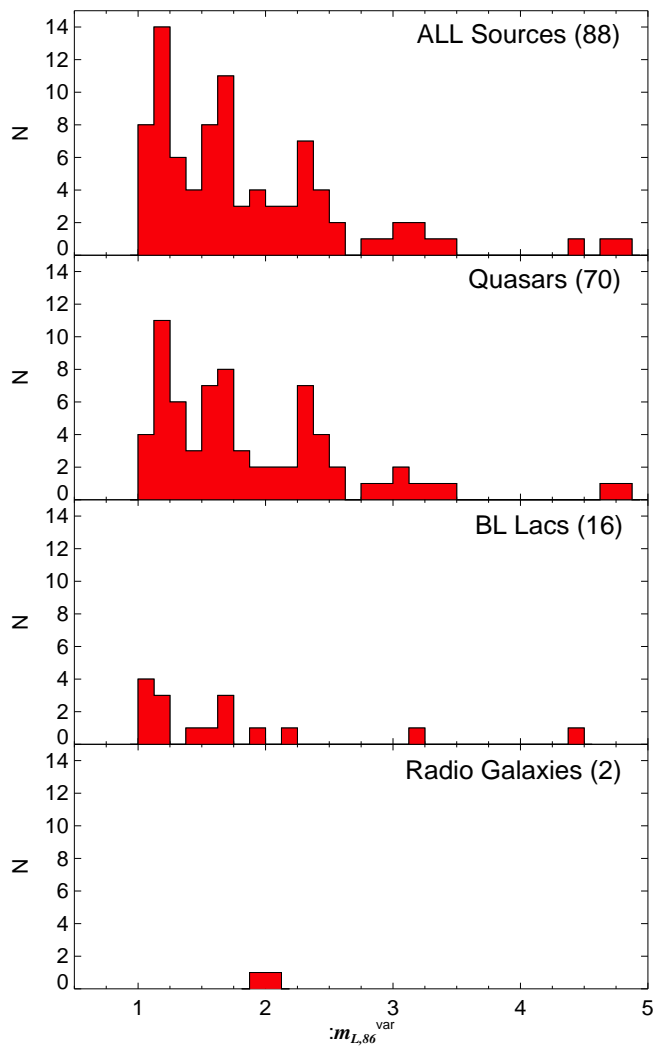}
   \caption{Distribution of 86\,GHz polarization degree variability ratio ($:m_{L,86}^{\rm{var}}$, see the text) for the entire source sample, quasars, BL~Lacs and radio galaxies.}
   \label{m_Lvar}
\end{figure}

The distribution of 86\,GHz polarization degree variability ratio ($:\,m_{L,86}^{\rm{var}}$)\footnote{$:\,m_{L,86}^{\rm{var}}=\frac{max(m_{L,86}^{\rm{Agudo\,et\,al.\,(2010)}},m_{L,86}^{\rm{this\,paper}})}{min(m_{L,86}^{\rm{Agudo\,et\,al.\,(2010)}},m_{L,86}^{\rm{this\,paper}}})$} is shown in Fig.~\ref{m_Lvar}.
This Figure also points out an even larger degree of variability of $m_{L,86}$ with median variability factor $:m_{L,86}^{\rm{var}}=1.7$, and 34\,\% of the sources displaying increased (or decreased) $m_{L,86}$ by a factor of 2 in time scales of years.
There is also no significant difference between the different subsamples considered in Fig.~\ref{m_Lvar}.

\begin{figure}
   \centering
   \includegraphics[width=8.5cm]{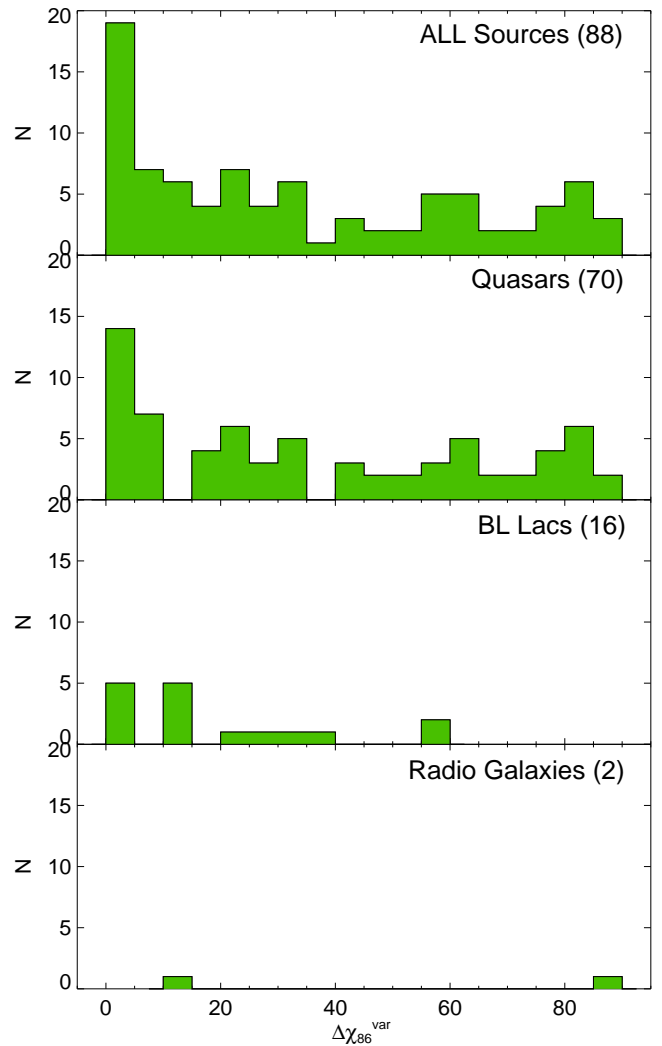}
   \caption{Distribution of the absolute difference of the 86\,GHz polarization angle from \citet{Agudo:2010p12104} and in this paper ($\Delta\chi_{86}^{\rm{var}}$, see the text) for the entire source sample, quasars, BL~Lacs and radio galaxies.}
   \label{chivar}
\end{figure}

The 86\,GHz linear polarization angle is also highly variable in time scales of years as shown by Fig.~\ref{chivar}, where we present the distribution of the absolute difference of the 86\,GHz polarization angle from \citet{Agudo:2010p12104} and in this paper ($\Delta\chi_{86}^{\rm{var}}$)\footnote{$\Delta\chi_{86}^{\rm{var}}=|\chi_{86}^{\rm{Agudo\,et\,al.\,(2010)}}-\chi_{86}^{\rm{this\,paper}}|$} for the entire source sample, quasars, BL~Lacs and radio galaxies.
Figure~\ref{chivar} also shows that the difference of 86\,GHz linear polarization angle of blazars in time scales of years is essentially evenly distributed among all possible angles between $0^{\circ}$ and  $90^{\circ}$, except for a small fraction (14\,\%) of sources that tend to conserve the same polarization angle in such time scales.
There is a 90.2\,\% probability for the quasar and the BL~Lac distributions in Fig.~\ref{chivar} to come from different parent distributions, but this is not enough for us to claim a statistically significant difference between them with regard to the behavior of their polarization angle variability.

Radio loud AGN, and blazars in particular, are able to show extreme total flux density variability in the millimeter range (and in other spectral ranges)  by up to one order of magnitude in time scales from months \citep{Jorstad:2005p264, Terasranta:2005p9128, Fuhrmann:2008p267} to days \citep{Agudo:2006p203}.
Such extreme variability (enhanced by relativistic Doppler boosting) is often connected to the ejection and propagation of strong jet perturbations (blobs or shocks) from the innermost regions of the source  \citep[e.g.,][]{Jorstad:2005p264, Kadler:2008p397, Perucho:2008p298,Jorstad:2010p11830,Marscher:2010p11374,Agudo:2011p14707,Agudo:2011p15946}.
Regarding linear polarization variability at short millimeter wavelengths, blazars have been shown to display $m_{\rm{L}}$ excursions of up to one order of magnitude and $\chi$ rotations by $>90^{\circ}$ in time scales of months or even weeks \citep[e.g.][]{Jorstad:2005p264,2007ApJ...659L.107D, Darcangelo:2009p6953,Jorstad:2010p11830,Agudo:2011p14707,Agudo:2011p15946}.
It is thus not surprising that a large level of variability affects our results, and of course the completeness of our sample (see  \S~\ref{Samp}).
The expected influence of such variability in our study is to broaden the $S$, $m_{\rm{L}}$, and $\chi$ distributions or to hide their correlation with other variables, hence making it more difficult to obtain statistically significant relations, but never faking results to give unrealistic correlations. 
We are therefore confident on the significance of the results shown here.

\section{Summary}

We have presented the first simultaneous 3.5 and 1.3\,mm polarization survey of radio loud AGN on a large sample of 211 sources, dominated by blazars.

Our total flux measurements show that almost all measured sources (96\,\%) have spectral index $\tilde\alpha_{86{\rm,}229}^{\rm{Q}}<0$ --most of them with optically thin spectrum-- between 86 and 229\,GHz with median spectral indexes  $\tilde\alpha_{86{\rm,}229}^{\rm{Q}}=-0.75$ for quasars and $\tilde\alpha_{86{\rm,}229}^{\rm{B}}=-0.56$ for BL~Lacs.
In contrast, the 15 to 86\,GHz spectral index is distributed towards flatter and more mildly optically-thin spectral-indices both for quasars and BL~Lacs.

We detect linear polarization above $3\sigma$ levels for 88\,\% and 13\,\% of the sources detected at 86 and 229\,GHz, respectively.
We find the distributions of ${m}_{\rm{L}}$ for quasars and BL~Lacs to be significantly different at both 86 and 229\,GHz with median linear polarization degrees $\tilde{m}_{\rm{L,86}}^{\rm{Q}}=3.2$\,\% and $\tilde{m}_{\rm{L,229}}^{\rm{Q}}=7.7$\,\% for quasars, and $\tilde{m}_{\rm{L,86}}^{\rm{B}}=4.6$\,\% and $\tilde{m}_{\rm{L,229}}^{\rm{B}}=10.2$\,\% for BL~Lacs.
The differences between quasars and BL~Lacs is explained by the fact that quasars have, in general, smaller viewing angles than BL Lacs, which produce stronger depolarization in quasars if the magnetic fields in the jets of blazars are not homogeneously distributed along the jet axis or the jets themselves are not axisymmetric.

We show for the first time that the 229\,GHz linear polarization degree is, in general, also a factor $\sim1.6$ larger than the one at 86\,GHz.
We also confirm that the 86\,GHz is in general stronger than that at 15\,GHz by a factor of $\sim1.6$.
This evidence implies that the magnetic field is progressively better ordered in blazar jet regions located progressively upstream in the jet.

An anti-correlation between the millimeter luminosity and the linear polarization degree is confirmed here for the entire source sample and quasars at 86\,GHz.
We attribute this relation to purely relativistic and orientation effects, i.e. sources whose jets are better oriented to the line of sight display larger luminosities (because of their larger Doppler boosting) and also lower linear polarization degrees (because of cancellation of orthogonal polarization components along the line of sight).

Unlike theoretical predictions from axially symmetric jet models, we show here an essentially inexistent relation between $\chi_{86}$ and the jet structural position angle for both quasars and BL~Lacs  to distribute at a preferential misalignment angle \citep[see also][]{Agudo:2010p12104}. 
Only a small 17\,\% of quasars tend to have $0^{\circ}\le|\chi_{86}-\phi_{\rm{jet}}|\lesssim30^{\circ}$.
This result seems to be reproduced also at 229\,GHz, although at this frequency the small number of sources detected in polarization do not allow to obtain robust conclusions. 
These results imply that either the magnetic field or the emitting particle distributions (or both) responsible for the synchrotron radiation in the innermost regions where the short mm emission is radiated in blazars have a markedly non axisymmetric character in general. 

Comparison of our data with the 86\,GHz measurements presented in \citet{Agudo:2010p12104} shows a considerably high total flux variability factor ($:S_{86}^{\rm{var}}$) with median variations $\bar{:S_{86}^{\rm{var}}}\sim1.5$ in time scales of years, with a 19\,\% of sources showing $:S_{86}^{\rm{var}}\gtrsim2$, and even one source (\object{0422+004}) experiencing  extreme total flux variations by $:S_{86}^{\rm{var}}=6.3$.
The 86\,GHz linear polarization degree is also equally highly variable, with median variability factor $\bar{:m_{L,86}^{\rm{var}}}\sim1.7$, and 34\,\% of sources displaying $:m_{L,86}^{\rm{var}}\gtrsim2$ with maximum variations by a factor of $\sim5$.
The 86\,GHz linear polarization angle is also highly variable in time scales of years.
Except for a small fraction (14\,\% of sources) of sources that tend to conserve the same polarization angle, most sources show drastically different polarization angle --evenly distributed among all possible angles between $0^{\circ}$ and $90^{\circ}$--  in such time scales.

\begin{acknowledgements}
     This paper is based on observations carried out with the IRAM 30~m Telescope. 
     IRAM is supported by INSU/CNRS (France), MPG (Germany) and IGN (Spain).
     The research at the IAA-CSIC is supported in part by the Ministerio de Ciencia e Innovaci\'{o}n of Spain, and by the regional government of Andaluc\'{i}a through grants AYA2010-14844 and P09-FQM-4784, respectively.
     The research at Boston University was funded by US National Science Foundation grant AST-0907893, NASA grants NNX08AJ64G, NNX08AU02G, NNX08AV61G, and NNX08AV65G, and NRAO award GSSP07-0009.
     This research has made use of the NASA/IPAC Extragalactic Database, the MOJAVE database \citep{Lister:2009p5316}, the one by the Blazar Research Group at the Boston University, as well as the USNO Radio Reference Frame Image Database.
\end{acknowledgements}




\end{document}